\documentclass[prl,aps,a4,epsfig,preprint]{revtex4}
\usepackage{subfigure}
\usepackage{amssymb}
\usepackage{amsmath}
\usepackage{epsfig}
\renewcommand{\dj}{Deutsch-Jozsa }
\newcommand{\qph}{ quant-ph/}
\newcommand{\qic}{ Quant. Inf. Comp }
\begin{document}
\large
\title{Experimental implementation of local adiabatic evolution algorithms by an NMR quantum information processor}
\author{Avik Mitra, Arindam Ghosh, Ranabir Das, Apoorva  Patel$^{\dagger}$ and Anil Kumar\footnote{DAE-BRNS,Senior scientist. {\it email}: anilnmr@physics.iisc.ernet.in}}
\affiliation{NMR Quantum Computation and Quantum Information group \\
Department of Physics and NMR Research Centre, \\ Indian Institute of Science, 
Bangalore - 560012, India\\
$^{\dagger}$Center for High Energy Physics, Indian Institute of Science, Bangalore - 560012, India.}

\begin{abstract}
Quantum adiabatic algorithm is a method of solving computational problems by evolving the ground state of a 
slowly varying Hamiltonian. The technique uses evolution of the ground state of a slowly varying Hamiltonian to 
reach the required output state. In some cases, such as the adiabatic versions of Grover's search algorithm and Deutsch-Jozsa algorithm,  
applying the global adiabatic evolution yields a complexity similar to their classical algorithms. However, using the local adiabatic 
evolution, the algorithms given by J. Roland and N. J. Cerf for Grover's search [ Phys. Rev. A. {\bf 65} 042308(2002)] and by Saurya Das, 
Randy Kobes and Gabor Kunstatter for the Deutsch-Jozsa algorithm [Phys. Rev. A. {\bf 65}, 062301 (2002)], yield a  complexity of order 
$\sqrt{N}$ (where N=2$^{\rm n}$ and n is the number of qubits). In this paper we report the experimental implementation of these local 
adiabatic evolution algorithms on a two qubit quantum information processor, by Nuclear Magnetic Resonance.
\end{abstract}
\maketitle
\vspace{0.5cm}

\section{1. Introduction}
Quantum algorithms provide elegant opportunities to  harness available quantum resources and perform certain computational 
tasks more efficiently than classical devices. The idea that a quantum computer
could simulate the physical behavior of a quantum system as well as perform computation, attracted immediate attention \cite{preskill,ss}.
The theory of such quantum computers is now well understood and several quantum algorithms like Deutsch-Jozsa (DJ) algorithm \cite{deu},
Grover's search algorithm \cite{grover}, Shor's prime factorization algorithm \cite{shor}, Hogg's algorithm \cite{hogg},Bernstein-Vazirani 
problem \cite{vazi} and quantum counting \cite{count1} have been developed . All these algorithms start from a well-defined initial state 
and perform computation by a sequence of reversible logic gates. After computation, the final state of the system gives the output. 
Various methods are being examined for building a quantum information processing (QIP) device which is coherent and unitary \cite{bou}.
Nuclear Magnetic Resonance has emerged as a leading candidate
for implementation of various quantum computational problems on a small number of qubits 
\cite{cory97,chuang97,cory98,djchu,djjo,grochu,grojo,ka1,ka,jcp,nat,ranapra2,ijqi,ranapra1,ranabirtomo}.

Quantum adiabatic algorithms provide an alternative method for computing \cite{ad1,ad2}. In this method the 
computation is done by evolving the system under a Hamiltonian for a given amount of time. 
 Such algorithms start from a suitable input ground state and by evolution under a slowly time-varying Hamiltonian, reach the 
desired output state. Quantum adiabatic algorithms have been efficiently applied to solve various optimization problems 
\cite{ad3,ad4,ad5,ad6}. Chuang {\it et al.} have demonstrated the implementation of a quantum adiabatic algorithm by solving the MAX-CUT \cite{garey} 
problem on a three qubit system by NMR \cite{chu} . In these algorithms, the condition for adiabaticity is fulfilled globally by using only
the minimum energy gap between the ground state and the first excited state for calculating the time of evolution. This method of evolution
is not efficient in  some cases such as adiabatic Grover's search algorithm and adiabatic \dj algorithm as they result in a complexity 
O(N) (N is the size of the data set), which is as good as their classical algorithms. However, these algorithms can be 
improved by application of local adiabatic evolution, where the adiabatic condition is fulfilled at each instant of time. This technique 
has been adopted theoretically by Roland and Cerf \cite{cerf} for the adiabatic Grover's search algorithm and by S. Das 
{\it et al.} for adiabatic \dj algorithm \cite{das} yielding a complexity O($\sqrt{N}$). Experimental implementation of 
adiabatic Grover's search algorithm based on the proposal of Roland and Cerf and adiabatic \dj algorithm of S. Das {\it et al.}, is 
reported here. Section 2 contains an introduction to adiabatic algorithms. Section 3 discusses the adiabatic version of the Grover's search
algorithm proposed by Roland and Cerf and its NMR implementation. Section 4 discusses the adiabatic \dj algorithm and its NMR 
implementation. Section 5 contains the experimental results, on a 2-qubit system,  for both these  algorithms. To the best of our knowledge 
this is the first experimental implementation of adiabatic Grover's search and adiabatic \dj algorithms.

\section{2. Adiabatic Algorithm} 
The adiabatic theorem of quantum mechanics states that when a system is evolved under a slowly time varying Hamiltonian, it stays 
in its instantaneous ground state \cite{me}. This fact is used in solving certain 
computational problems \cite{ad3,ad4,ad5,ad6}. The problem to be solved is encoded in a final Hamiltonian ($H_F$), 
whose ground state is not easy to find. Adiabatic algorithms start with the ground state of a beginning Hamiltonian 
($H_B$) which is easy to construct and whose ground state is also easy to prepare. The ground state of $H_B$, which is a superposition 
of all the eigenstates of H$_F$, is evolved under a time varying  Hamiltonian $H(s)$. $H(s)$ is a  linear interpolation of the beginning 
Hamiltonian $H_B$ and the final Hamiltonian $H_F$ such that
\begin{eqnarray}
H(s)= (1-s)H_B + s H_F, \hspace{3cm}\mbox{where}\;\;\; 0\leq s \leq 1. \label{hs}
\end{eqnarray}
The parameter $s=t/T_{total}$, where $T_{total}$ is the total time of evolution and $t$ varies from 0 to $T_{total}$. After evolution under
the Hamiltonian $H(s)$ for a time $T_{total}$, the system is in the ground state of $H_F$ with a probability $(1-\varepsilon^2)^2$, 
provided the evolution rate satisfies,
\begin{eqnarray}
\frac{\underset{0\leq s \leq 1}{max}\left|\left<1;s\left|\frac{dH(s)}{dt}\right| 0;s\right>\right|}{g_{min}^{2}} \leq \varepsilon, \label{r1}  \label{epsilon}
\end{eqnarray} 
and the parameters of the algorithm are chosen to make $\varepsilon  \ll$1 \cite{ad1}.
The numerator  in Eq. 2  is the transition amplitude between the ground state and the first excited state of {\it H(s)}, and the denominator
is the square of the smallest energy gap $(g_{min})$ between them. Ideally the time of evolution ($T_{total}$) must be infinite. However as 
long as the gap is finite, for any finite and positive $\varepsilon$, the time of evolution can be finite. The time of evolution 
of the algorithm is determined by the minimum energy gap between the ground state and the first excited state. 
In the adiabatic case the time of evolution determines the complexity of the algorithms (that is  how long
it takes for the task to be completed), which can then be compared to the complexity of the discrete algorithms in classical and 
quantum paradigms. The time of evolution is measured in units of natural time scale associated with the system, $\bar T$ which is 
O($\hbar /{\bar E}$) where $\bar E$ is the fundamental energy scale associated with the physical system used to construct the states.
 \cite{das}. 

\indent In the actual implementation, the Hamiltonian $H(s)$ is discretized into $M+1$ steps as 
$H(\frac{m}{M})$ where m goes from $0\rightarrow M$ \cite{chu,wvd}. Thus the time varying 
Hamiltonian $H(s)$ goes from beginning Hamiltonian to final Hamiltonian in M+1 steps. As the total number of steps increase, the evolution 
becomes more and more adiabatic \cite{chu}. The evolution operator for the m$^{\rm th}$ step is given by \cite{chu}
\begin{eqnarray}
U_m=e^{-i[(1-\frac{m}{M})H_B + \frac{m}{M}H_F]\Delta t},
\end{eqnarray}
where $ \Delta t=T/(M+1)$. The total evolution is given by,
\begin{eqnarray}
U=\prod_{m=0}^M U_m.
\end{eqnarray}
Since, $H_B$ and $H_F$ do not commute in general, the evolution operator of Eq. 3 is approximated to first order in $\Delta$t, by the use of
the  Trotter's formula \cite{chu} as
\begin{eqnarray}
U_m \approx e^{-iH_{B}(1-\frac{m}{M})\frac{\Delta t}{2}}\cdot e^{-iH_{F}\frac{m}{M}\Delta t}\cdot e^{-iH_{B}(1-\frac{m}{M})\frac{\Delta t}{2}}. \label{eq:trot}
\end{eqnarray}
Thus in each step only a small evolution of the system from ground state of ${\rm H_B}$ towards the ground state of ${\rm H_F}$ takes place.

\section{3. Grover's search algorithm}
Suppose we are given an unsorted database of N items and one of those items is marked. To search for the marked item classically, it would 
require on an average N/2 queries. However using quantum resources, the algorithm prescribed by Grover \cite{grover} performs the same 
search with O($\sqrt{N}$) queries. The algorithm starts with an equal superposition of states, representing the items, 
repeatedly flips the amplitude of the marked state (done by the oracle) followed by the flip of the 
amplitudes of all the states about the mean. The number of times this process is repeated determines the complexity of the algorithm and 
this scales with the size of the database as O($\sqrt{N}$). \\
\indent In the adiabatic version, the system is evolved under a time dependent Hamiltonian which is a linear interpolation of H$_B$ and  
H$_F$. As n qubits are used to label a database of size N (=2$^n$), the resulting Hilbert space is of dimension N. The basis
states in this space are $\vert i\rangle$ where i=0,$\cdots$,N. H$_B$ is chosen such that the ground state is a linear 
superposition of all the basis states. Therefore for a 2-qubit case,

\begin{eqnarray}
\vert\psi_B\rangle &=& \frac{1}{2}\left(\vert 00\rangle + \vert 01\rangle + \vert 10\rangle + \vert 11\rangle\right). \label{groinistate}\\
H_B &=& I - \vert\psi_B\rangle\langle\psi_B\vert, \cr
&=& I - \frac{1}{4}\begin{pmatrix}1&1&1&1 \cr 1&1&1&1 \cr 1&1&1&1 \cr 1&1&1&1\end{pmatrix}.\label{grohb}
\end{eqnarray}
The Final Hamiltonian has the marked state $\vert\psi_F \rangle$ as the ground state.
\begin{eqnarray}
H_F &=& I - \vert\psi_F \rangle\langle\psi_F \vert. \label{grohf}
\end{eqnarray}
The rate at which the interpolating  Hamiltonian H(s) (given by Eq. \ref{hs}) changes from $H_B$ to $H_F$ depends on the condition,
\begin{eqnarray}
\left|\frac{ds}{dt}\right| \leq \varepsilon \frac{g^{2}(s)}{\left|\langle\frac{dH}{ds}\rangle\right|}. \label{adcon}
\end{eqnarray}
Following Roland and Cerf \cite{cerf}, t is obtained as a function of s as,
\begin{eqnarray}
t=\frac{1}{2\varepsilon}\frac{N}{\sqrt{N-1}}\left[arctan\{\sqrt{N-1}\left(2s-1\right)\}+arctan\sqrt{N-1}\right].
\end{eqnarray}
Taking t$'= \varepsilon t$ and on inverting the above function, s(t$'$) is obtained as
\begin{eqnarray}
s(t') = \frac{1}{2}\left[\{ \frac{1}{\sqrt{N-1}}tan\left(\frac{2\sqrt{N-1}t'}{N} - arctan\sqrt{N-1} \right)\}+1\right]. \label{stprime}
\end{eqnarray}
The plot of this function for N=4 (for a 2 qubit case) is given in Fig. 1. In the experiment the time of
evolution is varied according to Eq. \ref{stprime}. It has been shown by Roland and Cerf \cite{cerf} that with this  
adiabatic evolution, the complexity of the algorithm is O($\sqrt{N}$). 

\section{3.1. Experimental Implementation}
The NMR Hamiltonian for a weakly coupled two-spin system is :
\begin{eqnarray}
{\mathcal H}= -\omega_1 I_{z1} - \omega_2 I_{z2} +2\pi J_{12}I_{z1}I_{z2}. \label{nmrham}
\end{eqnarray}
where $\omega_1$ and $\omega_2$ are Larmour frequencies and $J_{12}$ the indirect spin-spin coupling. The beginning Hamiltonian for a 
2-qubit Grover's algorithm as stated in Eq. \ref{grohb}, written in terms of spin-half operators, is
\begin{eqnarray}
{\mathcal H}_B = \frac{3}{4} I - \frac{1}{2} \{I_{x1} + I_{x2} + 2 I_{x1}I_{x2}\}. 
\end{eqnarray}
The identity term does not cause any evolution of the state and so it can be omitted, yielding the beginning Hamiltonian without 
the negative sign and the factor half as:
\begin{eqnarray}
\tilde{\mathcal H}_B =  I_{x1} + I_{x1} + 2I_{x1}I_{x2}  \label{nmrhb}
\end{eqnarray}
The evolution under $\tilde{\mathcal H}_B$ can be simulated by a free evolution under the Hamiltonian ${\mathcal H}$ of Eq. \ref{nmrham} between two
$\pi$/2 pulses with appropriate phases.
\begin{eqnarray}
e^{i\frac{\pi}{2}\left(I_{y1} + I_{y2}\right)}\cdot e^{i{\mathcal H}T}\cdot e^{-i\frac{\pi}{2}\left(I_{y1} + I_{y2}\right)}
&=&e^{i\left(\omega_1 I_{x1} + \omega_2 I_{x2} + 2JI_{x1}I_{x2}\right)T}\cr &=& e^{i{\mathcal H}'T} \label{hbevol}
\end{eqnarray}
Let the state $\vert 00\rangle$ be the marked state. The final Hamiltonian is,
\begin{eqnarray}
H_{F}^{\vert 00\rangle} = I - \begin{pmatrix}1&0&0&0\cr 0&0&0&0\cr 0&0&0&0\cr 0&0&0&0\cr\end{pmatrix}
\end{eqnarray}
In terms of spin operators the final Hamiltonian is,
\begin{eqnarray}
{\mathcal H}_{F}^{\vert 00\rangle} = \frac{3}{4} I - \frac{1}{2}\left[I_{z1} + I_{z2} + 2I_{z1}I_{z2}\right]
\end{eqnarray}
The final Hamiltonian keeping the spin operator terms only and without the negative sign and the factor half is
\begin{eqnarray}
\tilde{\mathcal H}_{F}^{\vert 00\rangle} = I_{z1} + I_{z2} + 2I_{z1}I_{z2} \label{grohf00}
\end{eqnarray}
Similarly the final Hamiltonian for other states being marked, in terms of the spin-half operators, is
\begin{eqnarray}
\tilde{\mathcal H}_{F}^{\vert 01\rangle}&=&I_{z1} - I_{z2} - 2I_{z1}I_{z2}  \label{grohf01} \\
\tilde{\mathcal H}_{F}^{\vert 10\rangle}&=&-I_{z1} + I_{z2} - 2I_{z1}I_{z2} \label{grohf10} \\
\tilde{\mathcal H}_{F}^{\vert 11\rangle}&=&-I_{z1} - I_{z2} + 2I_{z1}I_{z2} \label{grohf11}
\end{eqnarray}
\indent The schematic representation of the experiment  for the adiabatic Grover's algorithm in a two qubit system [consisting of a $^1$H 
spin and a $^{13}$C spin] is shown in Fig. 2a. The experiment is divided into three parts. The
first part (preparation part) consists of preparation of pseudo-pure state (PPS) followed by equal superposition. The second part is the
adiabatic evolution, and the third part is the tomography of the resultant state. The pulse programme for the
preparation of PPS and equal superposition is shown in Fig. 2b. The PPS is prepared by the method of spatial averaging \cite{du}. After
preparing PPS, equal superposition of states is obtained by application of the Hadamard gate on both the qubits. The Hadamard gate is
implemented by $(\pi/2)_y$ -pulses, followed by $\pi_x$ -pulses on both proton and carbon  spins (Fig. 2b) \cite{grochu}.
The next stage consists of adiabatic evolution which has been carried out in the present work in 60 steps. Each step of the 
adiabatic evolution (Figs. 2c, 2d, 2e and 2f) consists of evolution under the final Hamiltonian for a time $\tau$ sandwiched between two 
evolutions under the beginning Hamiltonian for a time (T-$\tau$)/2. T is the total evolution time for one step and 
is equal to 1/$\pi$J. The value of $\tau (= s\times \frac{1}{\pi J})$ varies from 0 to T takes place as `s' increases from 
0 to 1 according to Eq. \ref{stprime}, in 60 steps. The pulse sequence for the beginning Hamiltonian is a free  evolution of the system 
juxtaposed between two $\pi$/2 pulses with appropriate phases on each of the spins (the part marked as H$_B$ in Figs. 2c-2f). 
The pulse sequence for the final Hamiltonian depends on the marked state as stated in Eqs. \ref{grohf01}-\ref{grohf11}.                         
If the state $\vert 00\rangle$ is the marked state, then the pulse sequence for the implementation of the final Hamiltonian is a free 
evolution of the system under the NMR Hamiltonian juxtaposed between two $\pi$ pulses on each of the spins (Fig. 2c). Similarly, if the 
state $\vert 01\rangle$ is marked the pulse sequence for the final Hamiltonian is a free evolution of the system between two $\pi$ pulses 
on the spin 1 (Fig. 2d), if the state $\vert 10\rangle$ is marked then the pulse sequence is a free evolution between two $\pi$ pulses on 
the spin 2 (Fig. 2e) and if the state $\vert 11\rangle$ is marked, then the pulse sequence simulating the final Hamiltonian is just a free 
evolution of the system under the NMR Hamiltonian (Fig. 2f).

\indent The third stage of the experiment is the tomography of the final density matrix after the adiabatic evolution. The density matrix
of a 2-spin system is a 4$\times$4 matrix consisting of 6 independent off-diagonal complex elements (the remaining 6 are their complex
conjugates), and the four diagonal elements which are the populations of the various levels. The diagonal elements are measured by
90$^o$ pulses on each qubit preceded by a gradient pulse.
The six off-diagonal elements consist of four single quantum (SQ), one double quantum (DQ) and one zero quantum (ZQ) coherences. The real
and the imaginary SQ, DQ and ZQ coherences in terms of the spin operators are;
\begin{eqnarray}
{\rm SQ}^{real}_{i} &=& I_{ix}\pm 2(I_{ix}I_{jz}), \cr
{\rm SQ}^{imag}_{i} &=& I_{iy}\pm 2(I_{iy}I_{jz}), \cr
{\rm DQ}^{real} &=& 2(I_{ix}I_{jx} - I_{iy}I_{jy}), \cr
{\rm DQ}^{imag} &=& 2(I_{iy}I_{jx} + I_{ix}I_{jy}), \cr
{\rm ZQ}^{real} &=& 2(I_{ix}I_{jx} + I_{iy}I_{jy}), \cr
{\rm ZQ}^{imag} &=& 2(I_{iy}I_{jx} - I_{ix}I_{jy}),
\end{eqnarray}
where i$\neq$ j = 1,2 represents the qubits. Although the single quantum terms are directly observable, for proper scaling,
all the off-diagonal elements are observed by a common protocol of two experiments;
\begin{eqnarray}
{\rm A:}\hspace{3cm}\left(\frac{\pi}{2}\right)^{i}_{\phi_1}\left(\theta\right)^{j}_{\phi_2} \longrightarrow &G_{z}& \longrightarrow\left(\frac{\pi}{2}\right)^{i}_{y}, \\
{\rm B:}\hspace{3cm}\left(\frac{\pi}{2}\right)^{i}_{\phi_1}\left(\theta\right)^{j}_{\phi_2} \longrightarrow &G_{z}& \longrightarrow\left(\pi
\right)^{j}\left(\frac{\pi}{2}\right)^{i}_{y}.
\end{eqnarray}
where $\theta$ denotes the pulse angle, $\phi_1$, $\phi_2$ the pulse phases and $G_z$ a gradient pulse. The first two pulses of the
experiment A (depending on the pulse angle $\theta$ and the pulse phases $\phi_1$ and $\phi_2$) convert terms like
$I_{i\alpha}+2I_{i\alpha}I_{j\beta}$ into diagonal terms given by $I_{iz}+2I_{iz}I_{jz}$, where $\alpha$ and
$\beta$ denote the x, y, or z component of the spin operators of the first and the second qubit respectively. The gradient destroys all the
transverse magnetization retaining only the longitudinal terms. The last pulse converts the retained longitudinal magnetization
$I_{iz}+2I_{iz}I_{jz}$ into observable terms $I_{ix}+2I_{ix}I_{jz}$. Thus the magnitude of $I_{i\alpha}+2I_{i\alpha}I_{j\beta}$ is mapped on
to $I_{ix}+2I_{ix}I_{jz}$ which is then observed. In experiment B, a $\pi$-pulse is applied on the spin `j' just before the $\pi$/2 pulse on
the spin `i'. This creates the observable term $I_{ix}-2I_{ix}I_{jz}$. The sum and difference of the two experiments yields 2$I_{i\alpha}$
and 2$I_{i\alpha}I_{j\beta}$ respectively. Six different experiments are needed to be performed to map the whole density matrix (real and
imaginary). The various pulse angles and phases required during the experiment, and the resultant terms that are observed due to them are
given in Table I. Experiments I and II yield the SQ, and experiments III-VI yield the ZQ and DQ coherences.

\section{4. Deutsch-Jozsa Algorithm} 
The \dj Algorithm determines whether a binary function $f(x)$,
\begin{eqnarray*}
f(x\vert x\in\{ 0,1\}^n) \rightarrow \{0,1\},
\end{eqnarray*}
is Constant or Balanced \cite{dja}.A constant function implies that the function has the same value 0 or 1 for all $x$. A balanced 
function implies that the function {\it`f'} is 0 for half the values of $x$ and 1 for the other half . For a two qubit case the constant and
the balanced functions are given in Table II.

In the adiabatic version of the \dj algorithm, the beginning Hamiltonian and its ground state, for a two qubit system, is given by
Eq. \ref{grohb} and Eq. \ref{groinistate} respectively. The final Hamiltonian is given by Eq. \ref{grohf} and the  
ground state of the final Hamiltonian for two qubits is of the form \cite{das};
\begin{eqnarray}
\vert\psi_F\rangle &=& \alpha\vert00\rangle + \frac{\beta}{\sqrt 3}\left(\vert01\rangle + \vert10\rangle + \vert11\rangle \right), \label{djfinstate}
\end{eqnarray}
where
\begin{eqnarray}
\alpha &=& \frac{1}{4}\left|(-1)^{f(00)}+(-1)^{f(01)}+(-1)^{f(10)}+(-1)^{f(11)}\right|, \cr \beta^2 &=& 1-\alpha^2. \label{ab}
\end{eqnarray}
From Eq. \ref{ab}  it is seen that when $\alpha=1$ the function $f$ is constant, and when $\alpha=0$ then it 
is balanced. Thus $\alpha$ is chosen depending on whether the function to be encoded in the final Hamiltonian is constant or balanced.
\indent Using Eqs. \ref{hs},\ref{grohb},\ref{grohf},\ref{djfinstate} and \ref{ab} the matrix for the interpolating Hamiltonian($H(s)$) can be written as \cite{das};
\begin{eqnarray}
H(s)=I-\frac{1-s}{4}
\begin{pmatrix}
1 & 1 & 1 & 1 \\
1 & 1 & 1 & 1 \\
1 & 1 & 1 & 1 \\
1 & 1 & 1 & 1 
\end{pmatrix} -\frac{s}{3}
\begin{pmatrix}
3\alpha & 0 & 0 & 0 \\
0 & \beta & \beta & \beta \\
0 & \beta & \beta & \beta \\
0 & \beta & \beta & \beta 
\end{pmatrix}. \label{hsmatrix}
\end{eqnarray}
S. Das {\it et al.\;} have shown that on evolution under the Hamiltonian $H(s)$ takes the initial state $\vert\psi_B\rangle$ to the 
solution state $\vert\psi_{F}\rangle$ \cite{das}. In the next section we describe an NMR implementation of the above algorithm.

\section{4.1. NMR Implementation}
The adiabatic \dj algorithm also, is implemented on the 2-qubit system. The beginning Hamiltonian in terms of the spin-half 
operators is the same as given in Eq. \ref{nmrhb}, and its implementation has been discussed in section 3.1. \\
\indent The final Hamiltonian, obtained from Eqs. \ref{grohb}, \ref{grohf}, \ref{djfinstate} and \ref{ab}, for constant case ($\alpha$=1) yields,
\begin{align}
H_{F}^{c} = I - \begin{pmatrix}1&0&0&0 \cr 0&0&0&0 \cr 0&0&0&0 \cr 0&0&0&0\end{pmatrix},\\ 
\intertext{and for balanced case ($\alpha$=0) yields,}
H_{F}^{b} = I - \frac{1}{3}\begin{pmatrix}0&0&0&0 \cr 0&1&1&1 \cr 0&1&1&1 \cr 0&1&1&1\end{pmatrix}. 
\end{align}
The above final Hamiltonians in terms of spin-half operators can be written respectively as,
\begin{align}
{\mathcal H}_F^{c} = \frac{3}{4}I - \frac{1}{2}&(I_{z1} + I_{z2} + 2I_{z1}I_{z2}), \intertext{and,}
{\mathcal H}_F^{b} = \frac{3}{4}I -\frac{1}{3}&\biggl[-\frac{1}{2}(I_{z1} + I_{z2} + 2I_{z1}I_{z2}) + 2(I_{x1}I_{x2} + I_{y1}I_{y2}) \cr & + I_{x1} + I_{x2} - 2(I_{x1}I_{z2} + I_{z1}I_{x2})\biggr].
\end{align}
As the identity does not cause any evolution of the state we consider only the spin operator terms. Thus the final Hamiltonian keeping only
the spin operators (dropping the minus sign), for the constant case, can be written as
\begin{eqnarray}
\tilde{{\mathcal H}}^{c}_F &=& \frac{1}{2}\{I_{z1} + I_{z2} + 2I_{z1}I_{z2}\}, \label{nmrhcf}
\end{eqnarray}
and for the balanced case as 
\begin{eqnarray}
\tilde{{\mathcal H}}^{b}_F = - \frac{1}{6}(I_{z1} &+ I_{z2} + 2I_{z1}I_{z2})+\frac{2}{3}(I_{x1}I_{x2}+I_{y1}I_{y2}) \cr
                                                &+\frac{1}{3}I_{x1}+\frac{1}{3}I_{x2}-\frac{2}{3}(I_{x1}I_{z2}+I_{z1}I_{x2}). \label{nmrhbf}
\end{eqnarray}
The signs of Eqs. \ref{nmrhb}, \ref{nmrhcf} and \ref{nmrhbf} are changed for consistency.
Since the various terms in Eq. \ref{nmrhbf}  do not commute, the evolution under this Hamiltonian would require a complex pulse sequence in 
NMR. However, we have found that by keeping only the diagonal terms in the Eq. \ref{nmrhbf}, the pulse sequence simplifies considerably with
the information regarding the balanced nature of the problem still encoded in it. This truncated final Hamiltonian for the balanced case 
is given by;
\begin{eqnarray}
(\tilde{\mathcal H}^{b}_{F})^{trunc} = -\frac{1}{6}(I_{z1} + I_{z2} + 2I_{z1}I_{z2})  \label{nmrhbftrunc}
\end{eqnarray}
The opposite signs of Eq. \ref{nmrhcf} and Eq. \ref{nmrhbftrunc} distinguish the constant and the balanced case.
\indent In the following we show that the balanced nature of the \dj problem is still encoded in $(\tilde{\mathcal H}_{F}^{b})^{trunc}$. 
Substituting $\alpha=0$ and $\beta=1$ and dropping the off-diagonal terms from the last part of Eq. \ref{hsmatrix} , we obtain 
\begin{eqnarray}
\tilde{H}^{b}(s) =I-\frac{1-s}{4} \begin{pmatrix}
1&1&1&1\\1&1&1&1\\1&1&1&1\\1&1&1&1
\end{pmatrix}
-\frac{s}{3}\begin{pmatrix}
0&0&0&0\\0&1&0&0\\0&0&1&0\\0&0&0&1
\end{pmatrix}.
\end{eqnarray}
The eigenvalues of this Hamiltonian are:
\begin{eqnarray}
\lambda_0 &=&\frac{1}{6}\left[3+2s-\sqrt{9 + s(7s-15)}\right], \\
\lambda_1 &=&\frac{1}{6}\left[3+2s+\sqrt{9 + s(7s-15)}\right], \\
\lambda_2 &=&\lambda_3 = 1-\frac{s}{3}.
\end{eqnarray}
The values of $\lambda_0$, $\lambda_1$, $\lambda_2$ and  $\lambda_3$ as a function of `s' are plotted in Fig. 3.
$\lambda_0$ is the ground state.  As `s' increases from 0, $\lambda_0$ continues to be the ground state and becomes the ground state 
of the final Hamiltonian in the limit $s\rightarrow 1$. The eigenvectors corresponding to $\lambda_0$,
$\lambda_1$, $\lambda_2$ and $\lambda_3$ are respectively obtained as;
\begin{eqnarray}
v_0 \!=\! \begin{pmatrix}
\frac{3-s-2\sqrt{9-15s+7s^2}}{3(s-1)} \cr 1\cr 1\cr 1
\end{pmatrix},\;
v_1 \!=\! 
\begin{pmatrix}
\frac{3-s+2\sqrt{9-15s+7s^2}}{3(s-1)} \cr 1\cr 1\cr 1
\end{pmatrix},\;
v_2 \!=\!    
\begin{pmatrix}
0 \cr -1 \cr 0 \cr 1
\end{pmatrix},\;
v_3 \!=\!
\begin{pmatrix}
0 \cr -1 \cr 1 \cr 0
\end{pmatrix},
\end{eqnarray}
The final state to which the system converges after the evolution is
\begin{eqnarray}
\underset{s\rightarrow 1}{lim}\;v_0 = \begin{pmatrix}0\cr 1\cr 1\cr 1 \end{pmatrix},
\end{eqnarray}
which is the desired output state. 

The energy gap between the ground state and the states corresponding to 
$\lambda_2$ and $ \lambda_3$ goes to zero as $s\rightarrow 1$ as shown in Fig. 3. However, there is no transition from 
$\lambda_0$ to $\lambda_2$, $\lambda_3$ as the transition amplitude given by the numerator in 
Eq. \ref{epsilon} is zero in these cases. Therefore the transition amplitude from the ground state 
$\lambda_0$ to the next excited state $\lambda_1$ is relevant for calculation of $s(t)$. The minimum energy gap between $\lambda_0$ and
$\lambda_1$, needed in Eq. \ref{epsilon}, is obtained for $s \simeq 1$ as seen in Fig. 3. Since the algorithm
is implemented using local adiabatic evolutions we need to change $s(t)$ such that the adiabatic condition \cite{cerf}
\begin{eqnarray}
\frac{ds}{dt} \leq \varepsilon \frac{\left|g(s)\right|^2}{\left|\left<\frac{dH}{ds}\right>\right|},
\end{eqnarray}
is met at each time interval. Here g(s) is the energy gap between the ground state and the first excited state, given by
$\frac{1}{3}\sqrt{9 -15s + 7s^2}$ and $\left|\left< dH/ds\right>\right| = H_F - H_B$. The Hamiltonian is evolved at a rate that is a 
solution of 
\begin{eqnarray}
\frac{ds}{dt} = \varepsilon \frac{\left|g(s)\right|^2}{\left|H_F - H_B\right|} \label{dsdt}
\end{eqnarray}
On integrating Eq. \ref{dsdt}, we obtain {\it t} as a function of {\it s}. 
\begin{eqnarray}
t=\frac{1}{\varepsilon}\frac{14s-15}{2\sqrt{3}\sqrt{7s^2 -15s + 9}} + k,\label{eq:t}
\end{eqnarray}
where the constant of integration $k=\frac{5}{\varepsilon 2\sqrt 3}$ to obey $s=0$ at $t=0$.
Inverting this function we obtain  $s(t')$ as 
\begin{eqnarray}
%s(t')=\frac{3\left(15{t'}^2 + 3\sqrt{7{t'}^2 -3{t'}^4} -35\right)}{14\left(3{t'}^2 -7\right)}.
s(t')=\frac{3}{14}\left[ 5 - \frac{\sqrt{225 + 24t\left( 55\sqrt{3} - 183t +60\sqrt{3}t^2 - 18t^3\right)}}{3 + 20\sqrt{3}t - 12t^2}\right]
\end{eqnarray}
where $t'$ is $\varepsilon t$. 
The plot of $s$ as a function of $t'$ is shown in Fig. 4. From Figs. 3 and 4 it is seen that the rate of change of $s$ (and hence of
the Hamiltonian) is fast when the energy gap  between $\lambda_0$ and $\lambda_1$ is large, and slow when the gap is small. 
In practice the time of evolution for $H_B$ and $H_F$ is given by $(1-s)\times T$ and $s\times T$ respectively, 
where $T$ is 1/$\pi$J and $s$ is varied from 0 to 1 according to Eq. \ref{eq:t}.  In our implementation, the $t^{'}$ interval for which 
$s$ varies from 0 to 1 is divided in 80 equal steps, and the corresponding values of s for each step (calculated from Eq. 48) are 
substituted in the evolution time of $H_B$ and $H_F$. \\
\indent On integrating Eq. 46 from s=0 to s=1, we get the total time of evolution 
\begin{eqnarray}
T_{total}=\frac{1}{\varepsilon}\frac{2}{\sqrt{3}}{\bar T}. \label{ttotal}
\end{eqnarray}
T$_{total}$ is given in the units of ${\bar T}$ which is the time scale associated with the physical system used \cite{das}. 
The time scale associated with evolution under the NMR Hamiltonian is $\sim 10^{-3} s$. The total time of evolution of the experiment 
($T_{total}$) is given by 80$\times$T, where T is the time for one step (see Fig. 5b). 
For the choice $\varepsilon\sim 10^{-2}$, T $\sim 60\times 10^{-3}$s in our case.

\section{4.2. Experimental Implementation}
The experimental implementation of adiabatic \dj algorithm on a 2-qubit system [consisting of a $^1$H spin and a $^{13}$C spin] 
also consists of three parts namely preparation, adiabatic evolution and tomography of the final density matrix. The preparation of the 
pseudo pure state (PPS) and making of equal superposing of states as well as the tomography of the final states has already been discussed 
in section 3. So we only describe the method of implementation of the final Hamiltonian for the \dj algorithm.\\ 
\indent The pulse sequence for the implementation of the constant case final
Hamiltonian ($\tilde{\mathcal H}_{F}^{c}$) is given in Fig. 5b. The beginning Hamiltonian is implemented by a free evolution juxtaposed 
between $\pi$/2 pulses with required phases (Fig. 5b). The implementation of the final Hamiltonian for the constant case is a free evolution
under the NMR Hamiltonian of Eq. 14, juxtaposed between two $\pi$-pulses as shown in Fig. 5b. In the balanced case the implementation of the
beginning Hamiltonian is same as in Fig. 5b. However, the implementation of the final Hamiltonian $(\tilde{\mathcal H}_{F}^{b})^{trunc}$ is 
done in two parts [Fig. 5c]. The first part is a free evolution under the Hamiltonian given in 
Eq. 14 [${\rm T_f}$ period in Fig. 5c]. The operator corresponding to such an evolution for time $\tau$ will be of the form;
\begin{eqnarray}
e^{i\pi J(-I_{z1}-I_{z2}+2I_{z1}I_{z2})\tau}. \label{tf}
\end{eqnarray}
In the second evolution of $2\tau$, the chemical shifts are refocused so that the system evolves only under its scalar coupling Hamiltonian 
$2\pi JI_{z1}I_{z2}$. 
Just before and after the evolution  $\pi$-pulses with appropriate phases are put on each of the spins to flip the sign of the corresponding
spin operator [${\rm T_j}$ period in Fig. 5c]. The operator for the sequence of two pulses with an intermediate evolution for $2\tau$ 
is of the form
\begin{eqnarray}
e^{-i(I_{x1})\pi}\cdot e^{i\pi J(2I_{z1}I_{z2})2\tau}\cdot e^{i(I_{x1})\pi}= e^{-i\pi J(2I_{z1}I_{z2})2\tau}. \label{tj}
\end{eqnarray}
As these two evolutions given in Eq. \ref{tf} and Eq. \ref{tj} commute, the effective evolution for the 3$\tau$ period is:
\begin{eqnarray}
e^{i\pi J(-I_{z1}-I_{z2}+2I_{z1}I_{z2})\tau} \cdot e^{-i\pi J(2I_{z1}I_{z2})2\tau}  =e^{i\pi J(-I_{z1}-I_{z2}-2I_{z1}I_{z2})\tau}.\label{tftj}
\end{eqnarray}
Thus the evolution during ${\rm T_j}$ cancels the J-evolution during ${\rm T_f}$ and adds a minus sign to it, yielding the effective 
Hamiltonian of Eq. \ref{tftj} and an effective evolution time of $\tau$. An evolution time of $\tau$=1/$\pi$J implements the full 
Hamiltonian of Eq. 39 as required for adiabatic evolution. Overall the cycle time for each step for the balanced
case is increased to T$+2\tau$.\\

\section{5. Experimental Results}
The experiments have been carried out  using carbon-13 labeled chloroform ($\rm ^{13}CHCl_3$) where the two spins  $^1$H and $^{13}$C form
the two qubit system. The proton spin represents the first qubit and carbon-13 the second.  The sample of $\rm ^{13}CHCl_3$ was dissolved in
the solvent CDCl$_3$ and the experiments were performed at room temperature in a magnetic field of 11.2 Tesla. At this field the $^{1}\rm H$
resonance frequency is 500.13 MHz and the $^{13}\rm C$ resonance frequency is 125.76 MHz. During the 
entire experiment, the transmitter frequencies of $^{1}\rm H$ and $^{13}\rm C$ are set at a value $J/2$ away from resonance to achieve the 
condition $\omega_1=\omega_2=\pi$J. The equilibrium spectra of the two qubits are shown in Fig. 6a, and the spectrum corresponding to
$\vert 00\rangle$ PPS is shown in Fig. 6b.
To quantify the experimental result we calculate the {\it average absolute deviation} \cite{nures} of each element of the experimentally 
obtained density matrix from each element of the theoretically predicted density matrix given by,
\begin{equation}
\Delta x=\frac{1}{N^2}\sum^N_{i,j=1}\vert  x_{i,j}^{T} - x_{i,j}^{E} \vert  \label{error}
\end{equation}
where N=$2^n$ (n being the number of qubits), $x_{i,j}^T$ is $(i,j)^{th}$ element of the theoretically predicted density matrix and 
$x_{i,j}^E$ is $(i,j)^{th}$ element of the experimentally obtained density matrix.

\section{5.1. Grover's Search Algorithm}
The experimental spectra corresponding to the implementation of Grover's search algorithm on the above two qubit system are given
in Fig. 7. the spectra given in Figs. 7a(i-iv) contain the reading of populations after respectively searching states $\vert 00\rangle$,
$\vert 01\rangle$,$\vert 10\rangle$ and $\vert 11\rangle$. The population spectra are obtained by
application of a gradient followed by a $\pi$/2 pulse. Depending on the final state, the population spectra consist of one single spectral
line for each spin. These correspond to, $\vert 00\rangle$ $\rightarrow$ $\vert 01\rangle$  and $\vert 00\rangle$ $\rightarrow$ 
$\vert 10\rangle$transition when the searched state is $\vert 00\rangle$ (Fig. 7a-i); $\vert 01\rangle$ $\rightarrow$ $\vert 00\rangle$ and 
$\vert 01\rangle$ $\rightarrow$ $\vert 11\rangle$ when the search state $\vert 01 \rangle$ (Fig. 7a-ii); $\vert 10\rangle$ $\rightarrow$ 
$\vert 00\rangle$ and $\vert 10\rangle$ $\rightarrow$ $\vert 11\rangle$ when the search state $\vert 10 \rangle$ (Fig. 7a-iii); 
$\vert 11\rangle$ $\rightarrow$ $\vert 01\rangle$ and $\vert 11\rangle$ $\rightarrow$ $\vert 10\rangle$ when the search state 
$\vert 11 \rangle$ (7a-iv). The coherence spectra in Fig. 7b have been obtained by observing the searched state without application of 
any r.f. pulses. The absence of any signal in the spectra confirms that there is no single quantum coherences after the search. 
To check for the absence of zero quantum and double quantum coherences as well, the entire density matrix has been tomographed. 
Fig 8a shows the theoretical and the experimental density matrices after the adiabatic evolution, when state $\vert 00\rangle$ has been 
searched. The mean deviation of the experimentally obtained density matrix from the theoretically predicted one (calculated using 
Eq. \ref{error}) is 2.49$\%$. Similarly Figs. 8b, 8c and 8d contain the 
theoretically predicted and experimentally obtained density matrices when the states $\vert 01\rangle$, $\vert 10\rangle$ and 
$\vert 11\rangle$ have been searched. The mean deviation of the experimental density matrices from their theoretically predicted 
counterparts are 1.92$\%$, 1.89$\%$ and 1.97$\%$ respectively.

\section{5.2. Deutsch-Jozsa Algorithm}
{\bf{5.2.1 \em Constant case}}\\
\noindent For the constant case (Eq. \ref{ab}), the state expected after the evolution (using the pulse sequence 
given in Fig. 5b) is $\vert00\rangle$. The density matrix consists of population in $\vert 00\rangle$ state and no coherences. 
The spectrum corresponding to the population for such a state, obtained by application of a gradient followed by $\pi$/2 pulses on each 
of the spins, consists of one single quantum coherence in each spin (`Population spectrum' in Fig. 9a). The spectrum for coherence, 
observed without application of any pulses on any of the spins, has a near absence of any signal (`Coherence spectrum' in Fig. 9a).
Further confirmation of the final state is done by the tomography of the complete density matrix. The 
Fig. 10 shows the tomography  of the experimental and theoretically predicted density matrices of the final state for the constant case. 
The mean deviation of the experimental density matrix from the theoretical one is 5.28$\%$ \\ 
\indent{\bf{5.2.2 \em Balanced case}}\\
\indent 
For the balanced case [Eq. 31, $\alpha$=0 and $\beta$=1], the state expected after the evolution  (using the pulse sequence of Fig. 5c)
is  $\frac{1}{\sqrt{3}}\left(\vert01\rangle +\vert10\rangle +\vert11\rangle\right)$. The theoretical density matrix of the final state 
is given in Fig 11(a). This state theoretically has three diagonal elements, one SQ coherence of each qubit and a ZQ coherence between 
the two qubits, all of equal intensity. This state is confirmed by the spectra shown in Fig. 9b and the density matrix in Fig. 11(b). 
The mean deviation of the experimentally obtained density matrix from the theoretically predicted one is 17.2$\%$.
It is seen that in the density matrix obtained from experiment, the SQ coherence of $^{13}$C (second 
qubit) and the ZQ coherence between $^{13}$C and $^1$H have significantly reduced intensity, compared to the theoretically expected 
values. 

 There are three sources of error in adiabatic algorithms. $\varepsilon$ gives a measure of the first source of error. Theoretically 
the total time of evolution in adiabatic algorithms should be infinite. However, in practice the evolution is terminated once the 
state is supposed to have been reached with sufficiently high probability given by $(1-\varepsilon^2)^2$ which in our case (for 
$\varepsilon =10^{-2}$)is obtained to be 99.98$\%$. The second source of error is due to neglect of O($\Delta t^3$) terms in the 
Trotter's Formula (Eq. \ref{eq:trot}). The maximum error introduced due to this is $\approx$ 0.92 $\%$ which can be safely neglected.

The third source of error is due to decoherence effects arising from the interaction of the spins with their 
surroundings. To study decoherence, the relaxation times T$_1$ and T$_2$ of $^{1}$H and $^{13}$C were measured. The T$_{2}$ for SQ 
coherences were measured by CPMG sequence. For the measurement of ZQ and DQ coherence decay rate, the term I$_{1x}$I$_{2x}$ was created 
and its relaxation rate was measured by CPMG sequence. The T$_{2}$ of SQ coherence of $^{1}$H was found to be 3.4 s and for $^{13}$C it 
was found to be 0.29 s. The 
decay rate of I$_{1x}$I$_{2x}$ term was found to be 0.19 s. The T$_{1}$ for $^{1}$H and $^{13}$C measured from the initial part of the 
inversion recovery experiment was found to be 21 s for $^1$H and 16s for $^{13}$C. Using these measured values of T$_{1}$ and T$_{2}$ the 
simulation for the balanced case was repeated including relaxation using Bloch's equations \cite{ernst}. Significant decay of the carbon 
coherences was observed. The mean deviation of the of the experimental density matrix form the theoretical density matrix including
relaxation is found to be 8.0$\%$. \\
\indent The observed mean deviation between the theoretically expected and the experimentally obtained density matrices for the Grover's 
search and the constant case of the \dj are small ($<$ 2$\%$ and $<$ 6$\%$ respectively) while that for the balanced case of the \dj is 
large ($\sim$ 17$\%$). In the first two cases, the results are encoded in the diagonal elements of the density matrix, which are 
attenuated by the spin lattice relaxation, the times for which are large ($>$ 16 sec). On the other hand, in the balanced \dj case, 
there are off-diagonal elements as well which are attenuated by spin-spin relaxation, the times for which are small ($<$ 4 sec for $^1$H 
and $<$ 0.3 sec for $^{13}$C). The decoherence times thus have a large effect in this case. A correction for the decoherence has improved 
the mean deviation considerably (reduced to $\sim$ 8$\%$), confirming the succesful implementation of these algorithms.
\section{6. Conclusion} 
In this paper we have demonstrated the experimental implementation of Grover's search and \dj algorithms by using local adiabatic 
evolution in a two-qubit quantum computer by nuclear magnetic resonance technique. We have suggested a different Hamiltonian for the 
adiabatic \dj algorithm  which is diagonal in the computational basis and hence easier to implement by NMR. To the best of our knowledge 
this is the first experimental implementation of these two algorithms by adiabatic evolution.We believe that this work will provide 
impetus to solving other problems by adiabatic evolution.\\ 
\underline{Acknowledgment}: 
The authors thank K.V. Ramanathan for useful discussions. The use of DRX-500 NMR spectrometer funded by the Department of
Science and Technology (DST), New Delhi, at the NMR Research Centre (formerly Sophisticated Instruments Facility),
Indian Institute of Science, Bangalore, is gratefully acknowledged.
AK acknowledges ``DAE-BRNS" for senior scientist support and DST for a research grant for
``Quantum Computing by NMR".
%=====================================================================================

%\textheight=670pt

\newpage
\section{Table Captions}
\begin{itemize}
\item[Table I:] The phases $\phi_1$ and $\phi_2$ and the flip angle of the $\theta$ pulses of the experiments A and B (see Eq. 26 and 27).
 $\alpha$ and $\beta$ denotes the various components of the spin operator terms of the first and the second qubit whose 
magnitudes were determined by that particular experiment.
\item[Table II:] The two constant and the six balanced functions for the 2-bit \dj algorithm.
\end{itemize}
\newpage

\begin{center}
\begin{table}
\caption{}\label{t1}
\begin{tabular}{p{2cm}p{2cm}p{2cm}p{2cm}p{2cm}p{2cm}}
    & $\phi_1$&$\phi_2$ &$\theta$       &$\alpha$&$\beta$\\
I   &$\bar{Y}$& -       & 0             & X      & Z  \\
II  & X       & -       & 0             & Y      & Z \\ 
III &$\bar{Y}$&$\bar{Y}$&$\frac{\pi}{2}$& X      & X \\
IV  &$\bar{Y}$&  X      &$\frac{\pi}{2}$& X      & Y \\
V   & X       &$\bar{Y}$&$\frac{\pi}{2}$& Y      & X \\
VI  & X       & X       &$\frac{\pi}{2}$& Y      & Y 
%\tablecaption{Phases and angles required to perform the complete set of experiments for tomography of the density matrix}
\end{tabular}
\end{table}
\end{center}

\begin{center}
\begin{table}
\caption{}\label{t1}
\begin{tabular}{|c|c|c|}\hline
     &  \quad {\bf Constant} \quad & \quad {\bf Balanced}  \cr\hline
\quad f(00) \quad &\quad $0 \hspace{1cm} 1$ &\quad$ 1 \hspace{1cm} 1 \hspace{1cm} 1 \hspace{1cm} 0 \hspace{1cm} 0 \hspace{1cm} 0 \quad $\cr
\quad f(01) \quad &\quad $0 \hspace{1cm} 1$ &\quad$ 1 \hspace{1cm} 0 \hspace{1cm} 0 \hspace{1cm} 1 \hspace{1cm} 0 \hspace{1cm} 1 \quad $\cr
\quad f(10) \quad &\quad $0 \hspace{1cm} 1$ &\quad$ 0 \hspace{1cm} 1 \hspace{1cm} 0 \hspace{1cm} 0 \hspace{1cm} 1 \hspace{1cm} 1 \quad $\cr
\quad f(11) \quad &\quad $0 \hspace{1cm} 1$ &\quad$ 0 \hspace{1cm} 0 \hspace{1cm} 1 \hspace{1cm} 1 \hspace{1cm} 1 \hspace{1cm} 0 \quad $\cr
\hline
\end{tabular}
\end{table}
\end{center}

\newpage
\section{Figure Captions}
\begin{itemize}
\item[Fig 1:] The plot of $s$ as a function of t$'$ for the 2-qubit adiabatic Grover's search algorithm (ref. Eq. 13). s(t$'$)
is plotted for that interval of t$'$ in which `s' goes from 0 $\rightarrow$ 1.

\item[Fig 2:] Pulse sequence for the implementation of adiabatic Grover's search algorithm. The narrow filled pulses,
which donot have any angle specified on them, represents 90$^o$ pulses while the broad unfilled pulses represents 180$^o$ pulses. The 
phases ($\bar{\rm X}$ and $\bar{\rm Y}$ represents -X and -Y phases respectively) are specified on each pulse. (a) Schematic representation 
of the pulse programme. The part `preparation' consists of creation of PPS, followed by equal superposition. The second part is adiabatic 
evolution done in 60 iterations and the third part is tomography of the final density matrix. (b) Pulse sequence for preparation of PPS 
and equal superposition. (c) Pulse sequence for the adiabatic evolution when the $\vert 00\rangle$ is been searched. In this sequence the
system is evolved under H$_B$ for time T-$\tau$ and under H$_F$ for a time $\tau$. This is  repeated 60 times for various $\tau$ varying it
from 0 $\rightarrow$ T as s is varied from 0 $\rightarrow$ 1 according to the Eq. 13 in equal intervals of t$'$. The total time in 
each iteration is T (= 1/$\pi$J) which is 1.52 ms. The pulse sequences (d),(e) and (f) when H$_F$ is encoded to search for the states 
$\vert 01\rangle$, $\vert 10\rangle$ and $\vert 11\rangle$ respectively.

\item[Fig 3:] The eigenvalues of $\tilde{H}^{b}(s)$ for the \dj algorithm (Eqs. 41-43) plotted as a function of parameter $s$. 
$\lambda_0$ is the ground state. $\lambda_1$,$\lambda_2$ and $\lambda_3$ are the excited states. $\lambda_2$ and $\lambda_3$ are 
degenerate for all values of $s$. $\lambda_0$ approaches $\lambda_2$ and $\lambda_3$ as s$\rightarrow$1. $\lambda_1$ changes marginally 
as a function of $s$.

\item[Fig 4:] Plot of the parameter $s$ as a function of $t'$ for the \dj algorithm (Eq. 48). $s$ is  0 for $t'$=0 and $s$ is 1 for 
$t'=\frac{2}{\sqrt{3}}$.This shows that $s$ changes rapidly at the beginning when $\vert\lambda_0 - \lambda_1\vert$ is large (Fig. 1) 
and later it changes slowly as $\vert\lambda_0 -\lambda_1\vert$ becomes small.

\item[Fig 5:] Pulse programme for the implementation of the adiabatic \dj algorithm.
The narrow filled pulses which do not have any angle specified on them are 90$^o$ pulses and all the broad unfilled pulses are 180$^o$
pulses.The phase of the pulses are specified on each of them. The frequency offset is set at a value J/2 during all the evolutions. 
(a) Block diagram representation of the pulse programme. The preparation sequence has already been explained in Fig. 2. The 
adiabatic evolution is shown in (b) and (c). The measurement process is the tomography of the final density matrix which is explained 
in the text. 
(b)Pulse sequence of the adiabatic evolution under the interpolating Hamiltonian H(s) for the 
constant case. ${\rm H_B}$ represents the beginning Hamiltonian and the pulse sequence implements the evolution as given in Eq. 
\ref{hbevol}. 
${\rm H^{c}_{F}}$ represents the final Hamiltonian when the constant case in encoded in it. T is the effective time of evolution in each 
cycle and $\tau$ goes from $0\rightarrow {\rm T}$ slowly in 80 steps. (c) The pulse sequence of H(s) for the balanced case. 
${\rm H^{b}_{F}}$ represents the pulse sequence for the implementation of the final Hamiltonian when the balanced case is encoded in it. 
During the period ${\rm T_{f}}$, the evolution takes place under the free Hamiltonian given by Eq. \ref{nmrham}. During the period 
${\rm T_{j}}$, the evolution takes place under the J-coupling Hamiltonian given in Eq. \ref{ttotal}. The $\pi$-pulse between ${\rm T_f}$ 
and ${\rm T_j}$ and at the end of the ${\rm T_j}$ restores the correct sign of the coupling Hamiltonian such that the total evolution 
for 3$\tau$ is given by  and the effective time of evolution for ${\rm H^{b}_{F}}$ is just $\tau$. The time $\tau$ is incremented in 80
steps from 0 $\rightarrow$ T, where T (=1/$\pi$J) is 1.5 ms in our experiments.

\item[Fig 6:] 
(a) Equilibrium spectra of  $^{13}$CHCl$_3$. The small line in the middle of the $^1$H spectrum is due to proton of
unlabeled chloroform and the three small equal intensity lines in the $^{13}$C spectrum are due to the J-coupling of the deuteron with the
natural abundant $^{13}$C in the solvent CDCl$_3$
(b)The spectra corresponding to $\vert 00\rangle$ PPS. A single line with positive intensity on each of the spins confirm that only 
$\vert 00\rangle$ level is populated.

\item[Fig 7:] Results of Grover's search algorithm. 
(a) The population spectra of the final state after the search have been performed. The populations have been observed by a applying
a gradient followed by a $\pi$/2 pulse. Figs. 7a(i - iv) contain the population spectra respectively when the states $\vert 00\rangle$, 
$\vert 01\rangle$, $\vert 10\rangle$ and $\vert 11\rangle$ have been searched. 
(b) The spectra of coherences (obtained by observing without the application of any pulse) of the final states after the adiabatic 
evolution. Figs. 7b(i - iv) contain the coherence spectra when the search states are $\vert 00\rangle$, $\vert 01\rangle$,
$\vert 10\rangle$ and $\vert 11\rangle$ respectively.

\item[Fig 8:] The tomography of the real and imaginary parts of theoretically expected and experimentally obtained  density matrices 
for the search states in Grover's algorithm. 
(a) $\vert 01\rangle$ (b) $\vert01\rangle$ (c) $\vert01\rangle$ and (d) $\vert01\rangle$. The density matrices consist of just a real 
term on the diagonal corresponding to the population of the state that has been searched. 

\item[Fig 9:] 
(a) Spectra of the population and coherence of $^1$H and $^{13}$C after the implementation of the constant case of the \dj algorithm. As the
final state for the constant case is $\vert 00\rangle$, the `Population' spectrum consists of one single quantum coherence for each spin 
and the `Coherence' spectrum  contains no signal [see text].
(b) The spectra of population and coherence for the balanced case of the \dj algorithm. The expected state is $\frac{1}{\sqrt{3}}
(\vert01\rangle+\vert10\rangle+\vert11\rangle)$. As the population of $\vert01\rangle$, $\vert10\rangle$ and $\vert11\rangle$ states are 
the same, the spectra  consists one SQ coherence for each of the spins. The coherence spectrum consists of another observable SQ 
coherence for each spin [see text]. The observed intensity of the SQ coherences in the `Population spectrum' is nearly half compared to 
those in the `Coherence spectrum' according to the expectations.

\item[Fig 10:] The tomography of the real and the imaginary parts of (a) theoretically expected and (b) experimentally obtained density 
matrices after the implementation of the constant case of the adiabatic \dj algorithm. As the theoretically expected state is 
$\vert 00\rangle$, the density matrices contains just one diagonal term which is the population of the $\vert 00\rangle$ state.

\item[Fig 11:] The tomography of the real and the imaginary parts of the (a) theoretically predicted and 
(b) experimentally obtained density matrices of the final state for the balanced case of adiabatic \dj algorithm. $^1$H and $^{13}$C are 
taken as the first and second qubit respectively. The theoretically predicted final state is $\frac{1}{\sqrt{3}}($$\vert 01\rangle$+$\vert 
10\rangle$+$\vert 11\rangle)$. Therefore the density matrix contains three diagonal elements corresponding to the populations of $\vert 
01\rangle$, $\vert 10\rangle$ and $\vert 11\rangle$ states, SQ coherences corresponding to $^1$H and $^{13}$C, and one ZQ coherence 
between the two qubits, all of equal intensity. 
(c) The real and the imaginary parts of the theoretically calculated density matrix of the final state for the balanced case of the \dj 
algorithm after the inclusion of the relaxation effects. The decay rates used for the single quantum coherences are 3.4s for $^1$H 
and 0.29 s for $^{13}$C. The decay rate of the zero quantum and double quantum coherences of this hetronuclear two spin system used 
is 0.19 s. 
\end{itemize}

\newpage
\begin{figure}[h!]
\epsfig{file=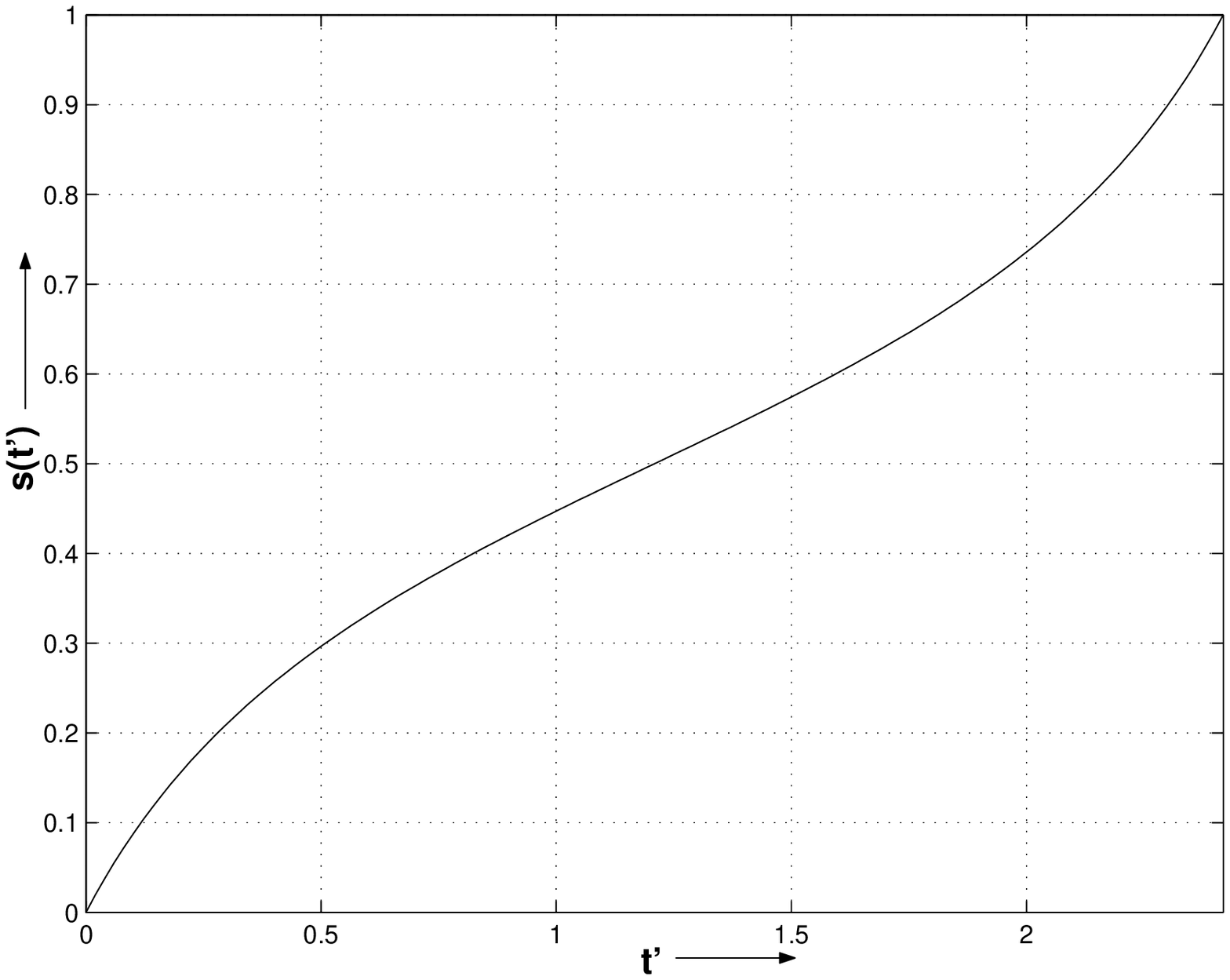,width=0.85\textwidth}
\caption{}
\end{figure}

\newpage
\begin{figure}[h!]
{\bf (a)} \raisebox{-12ex}{\epsfig{file=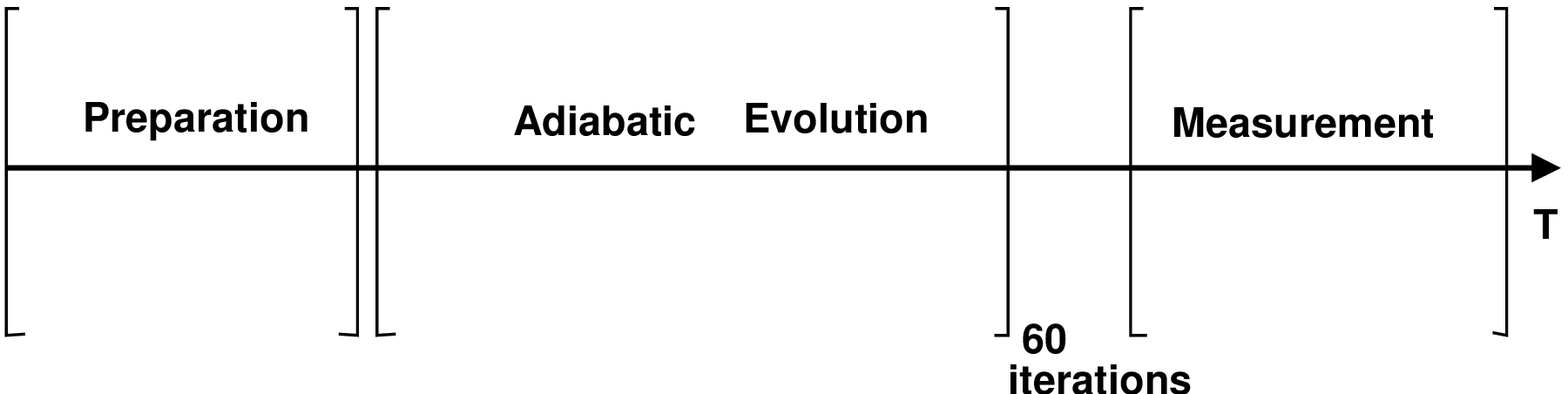,width=0.60\textwidth}} \\
{\bf (b)} \raisebox{-20ex}{\epsfig{file=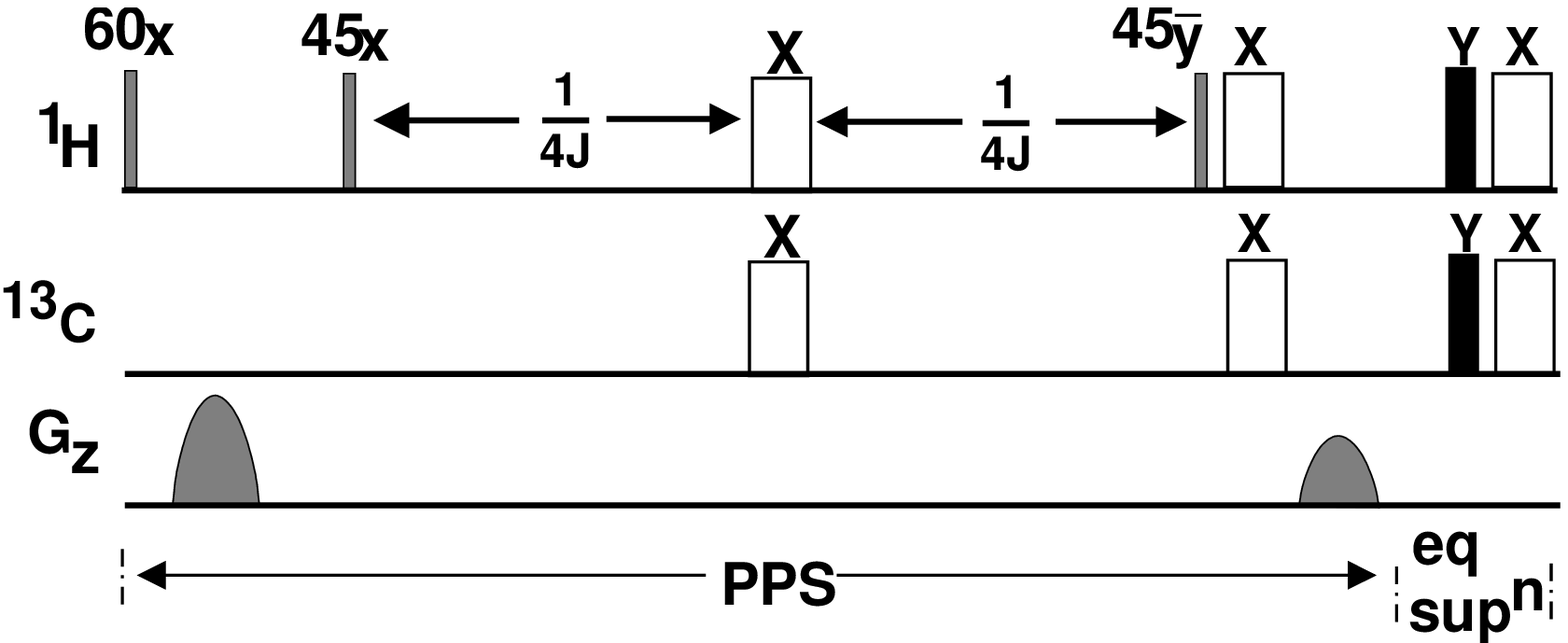,width=0.60\textwidth}} \\
{\bf (c)} \raisebox{-17ex}{\epsfig{file=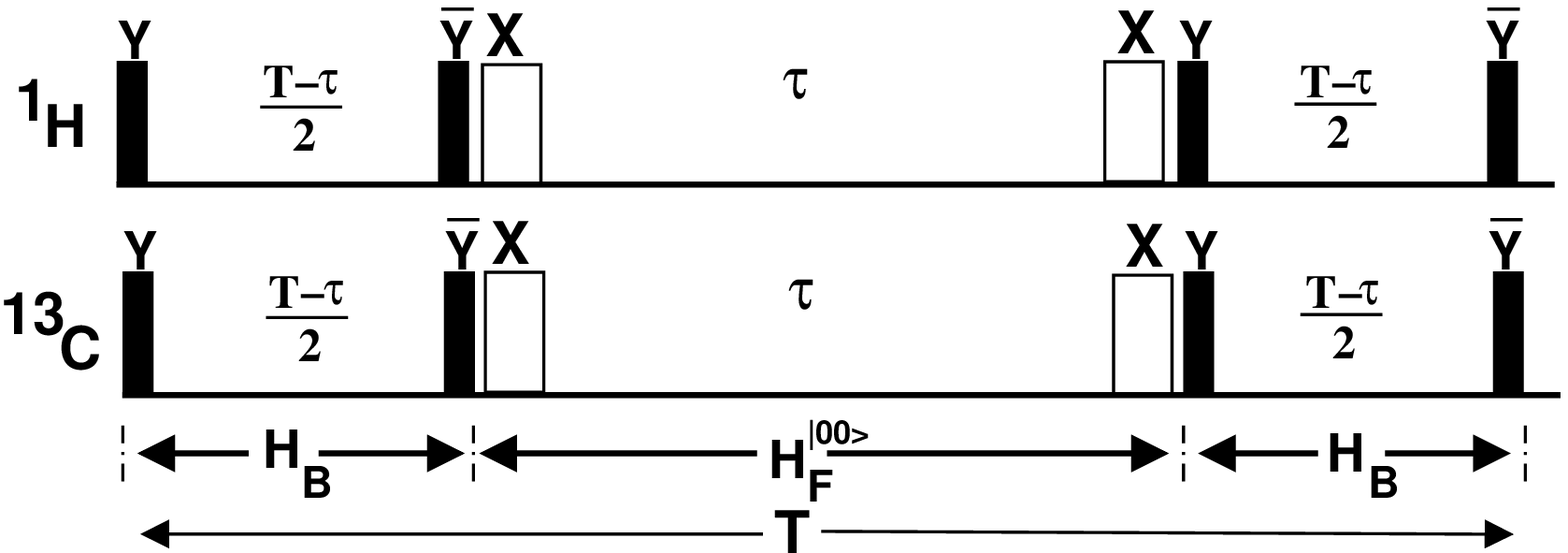,width=0.60\textwidth}} \\
{\bf (d)} \raisebox{-17ex}{\epsfig{file=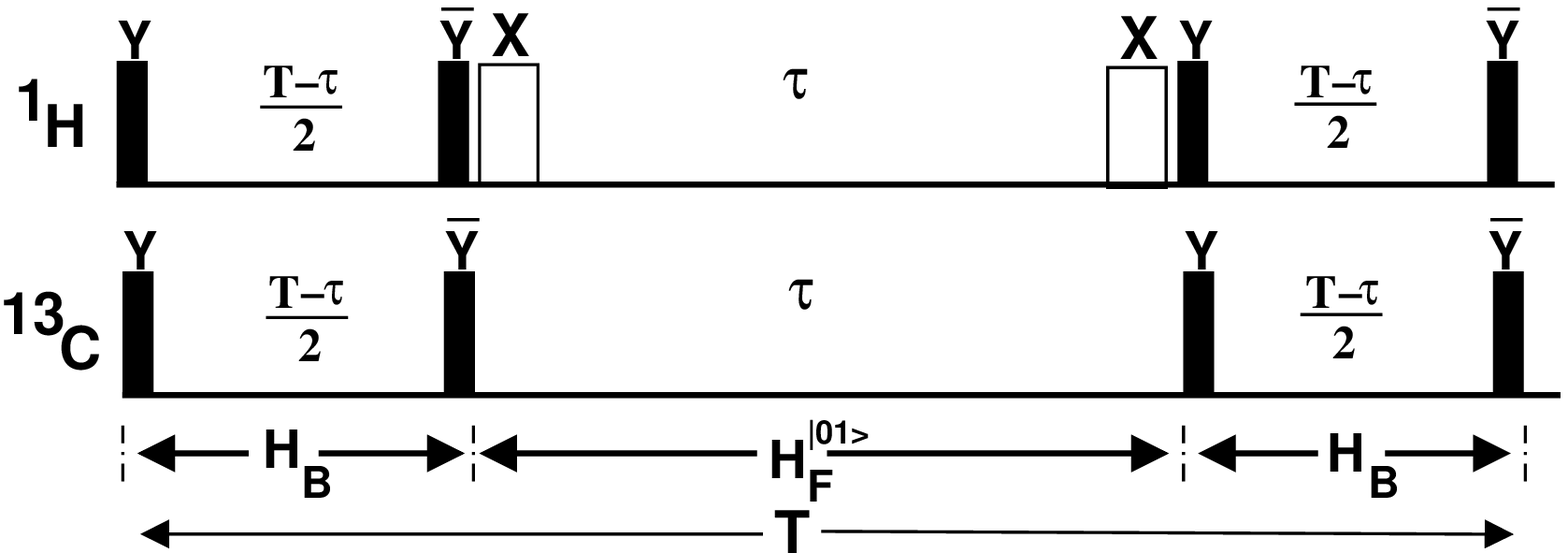,width=0.60\textwidth}} \\
{\bf (e)} \raisebox{-17ex}{\epsfig{file=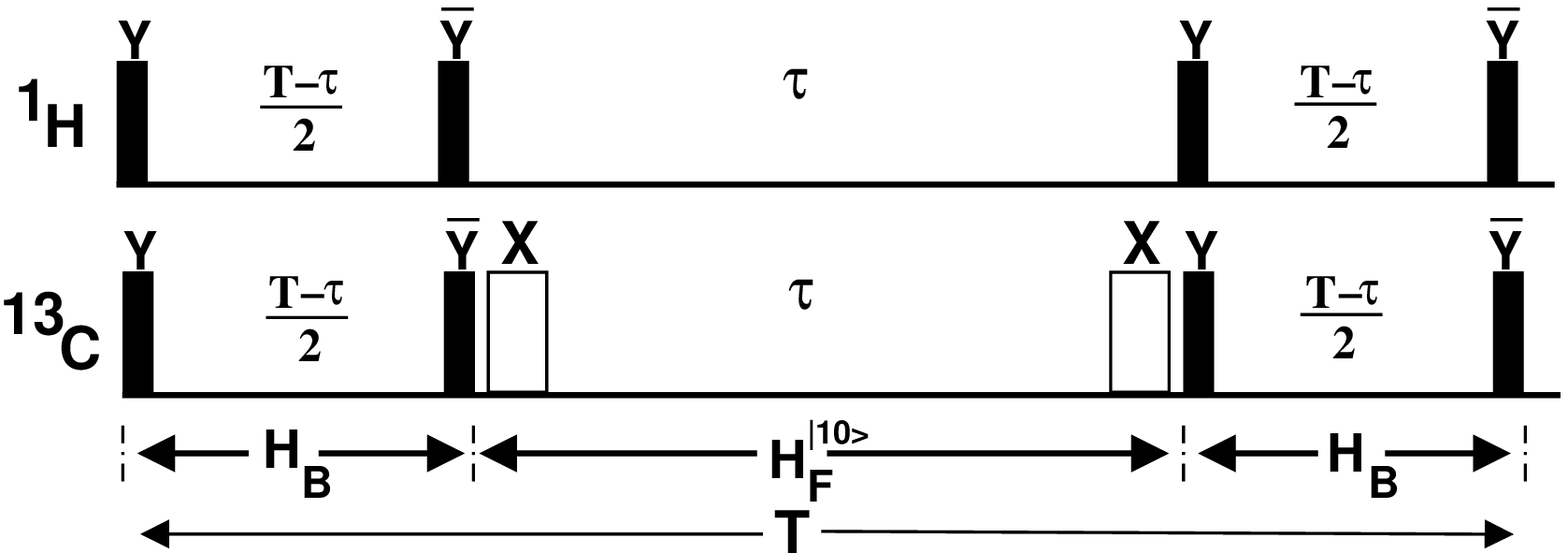,width=0.60\textwidth}} \\
{\bf (f)} \raisebox{-17ex}{\epsfig{file=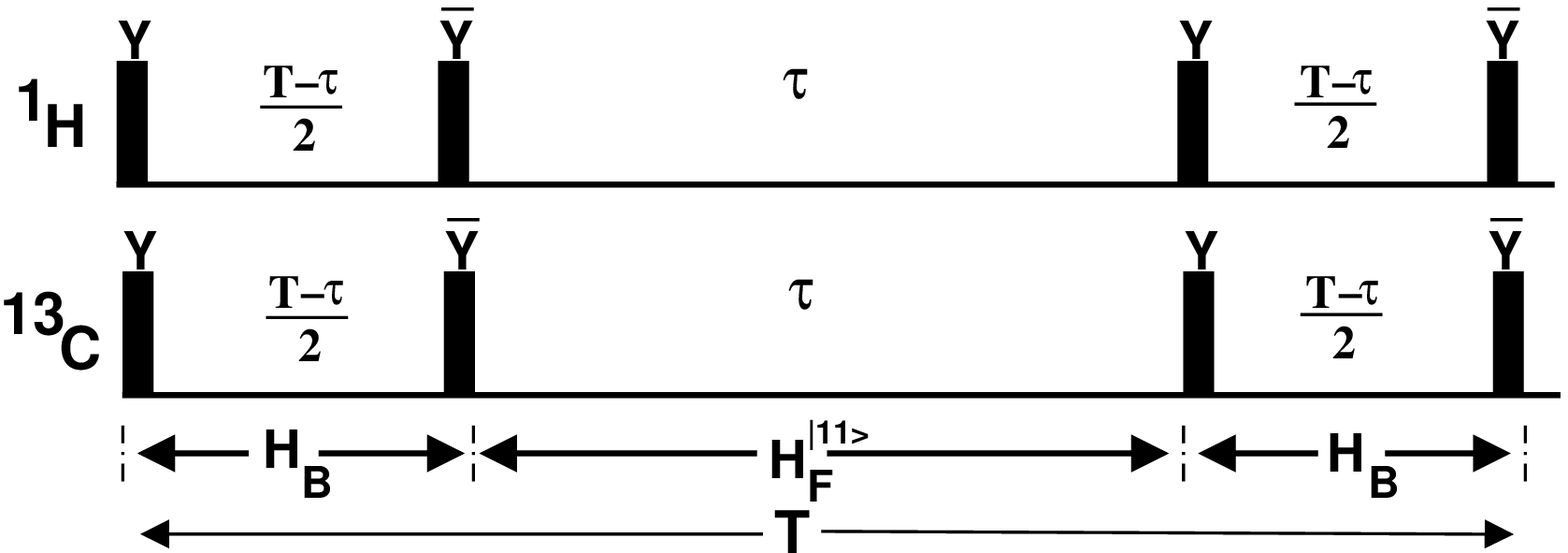,width=0.60\textwidth}} 
\caption{}
\end{figure}

\newpage
\begin{figure}[h!]
\epsfig{file=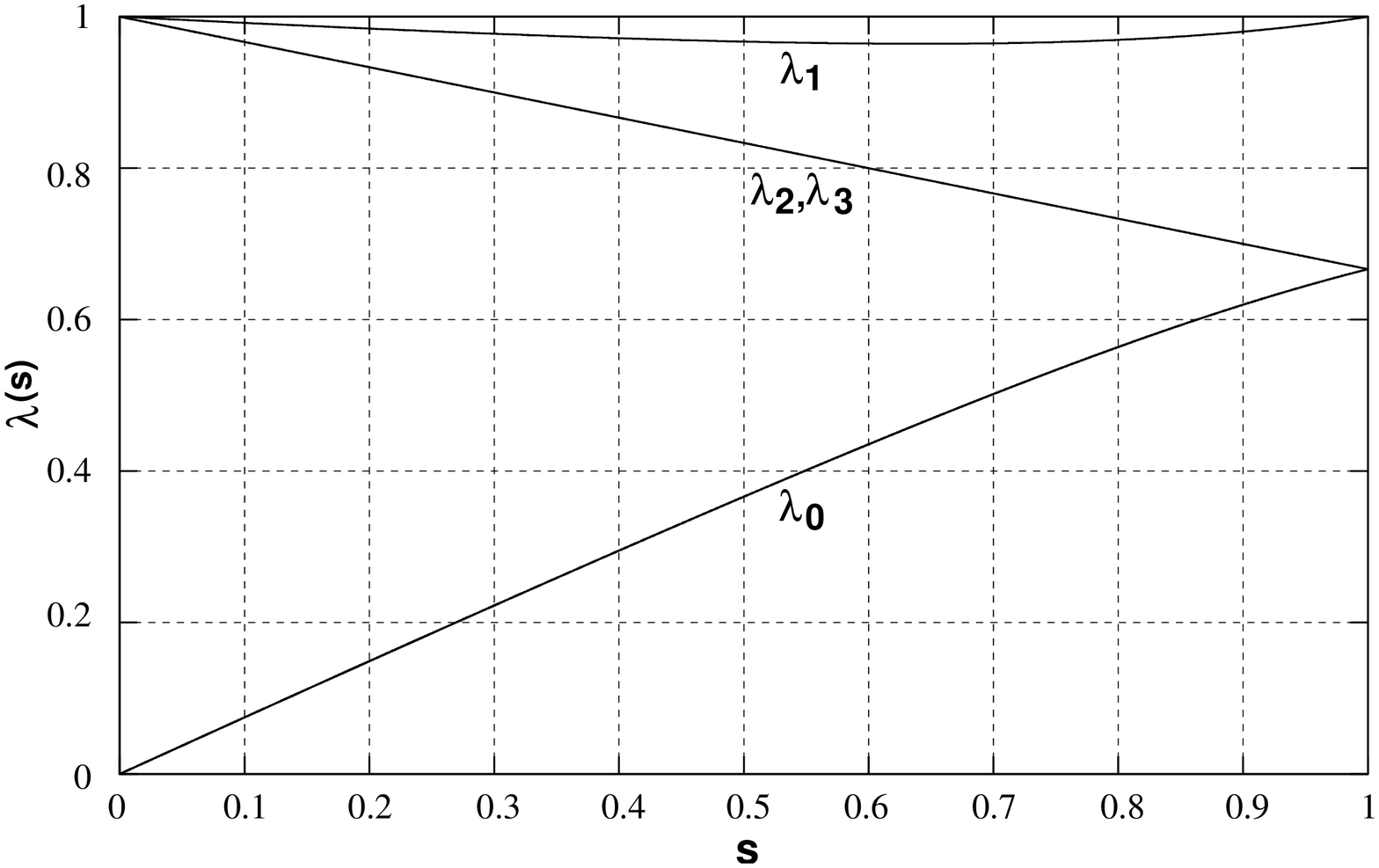,width=0.85\textwidth}
\caption{}
\end{figure}

\newpage
\begin{center}
\begin{figure}[h!]
\epsfig{file=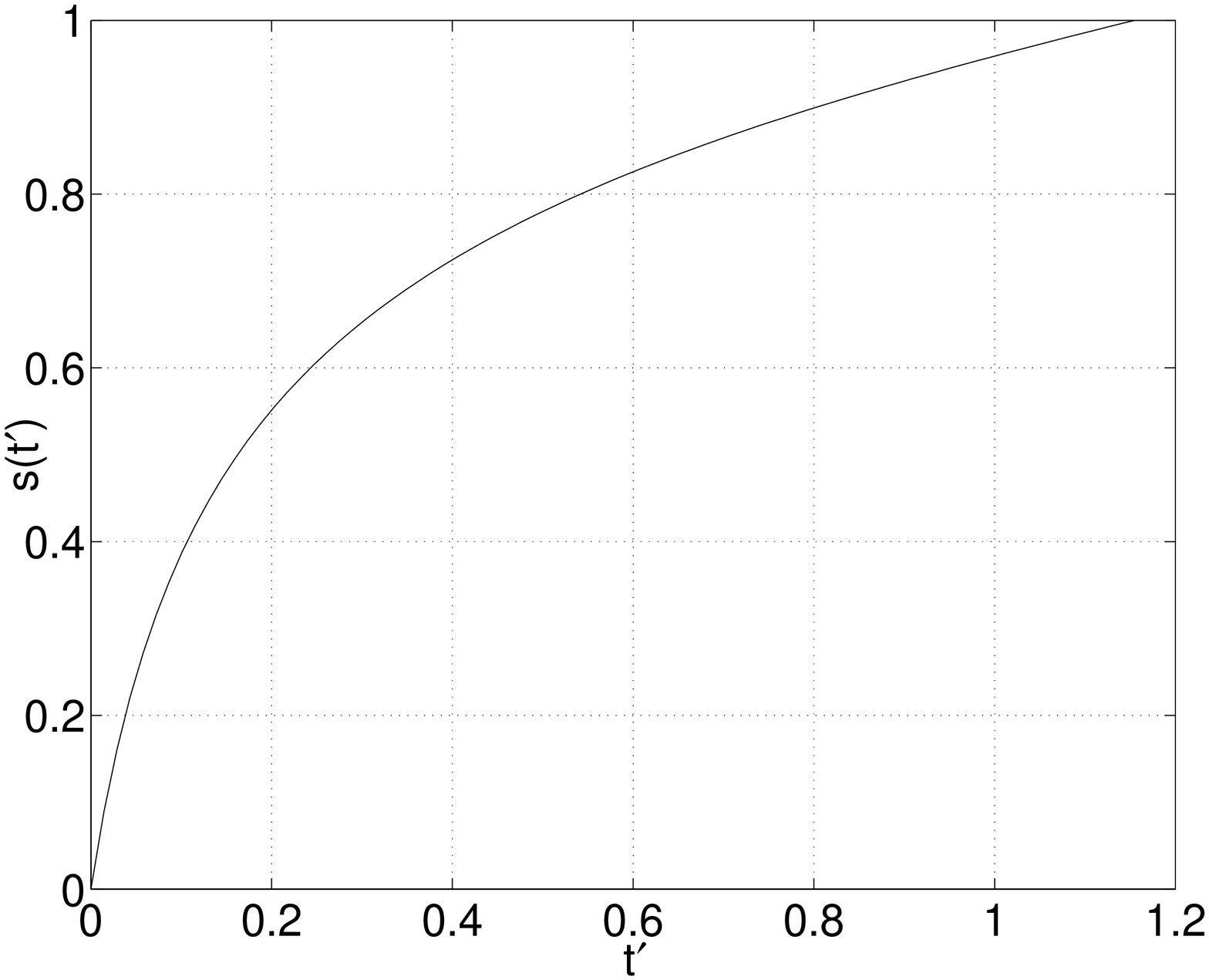,width=0.65\textwidth,angle=270}
\caption{}
\end{figure}
\end{center}

\newpage
\begin{center}
\begin{figure}[h!]
\subfigure{\epsfig{file=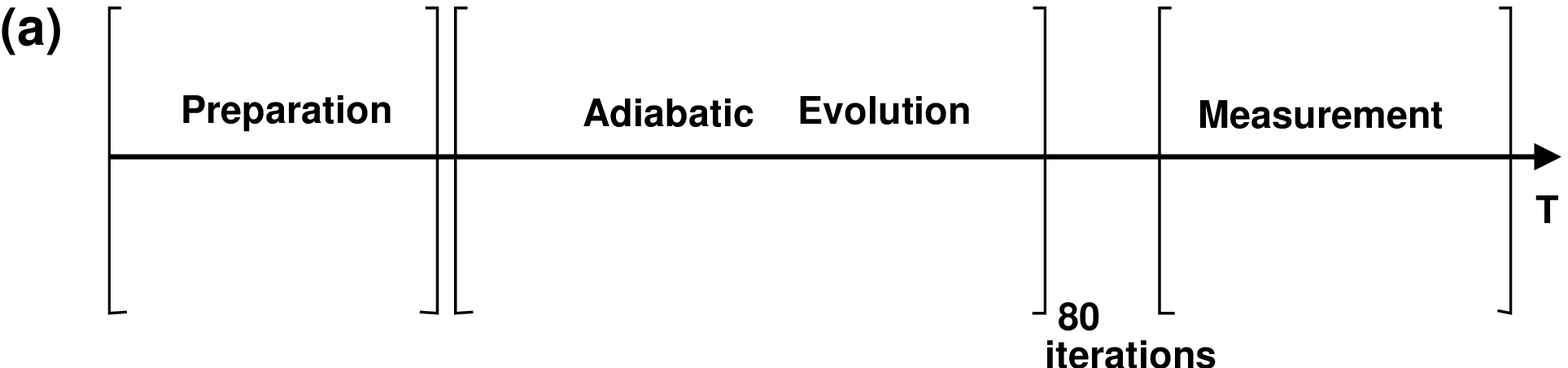,width=0.75\textwidth}}\\
\subfigure{\epsfig{file=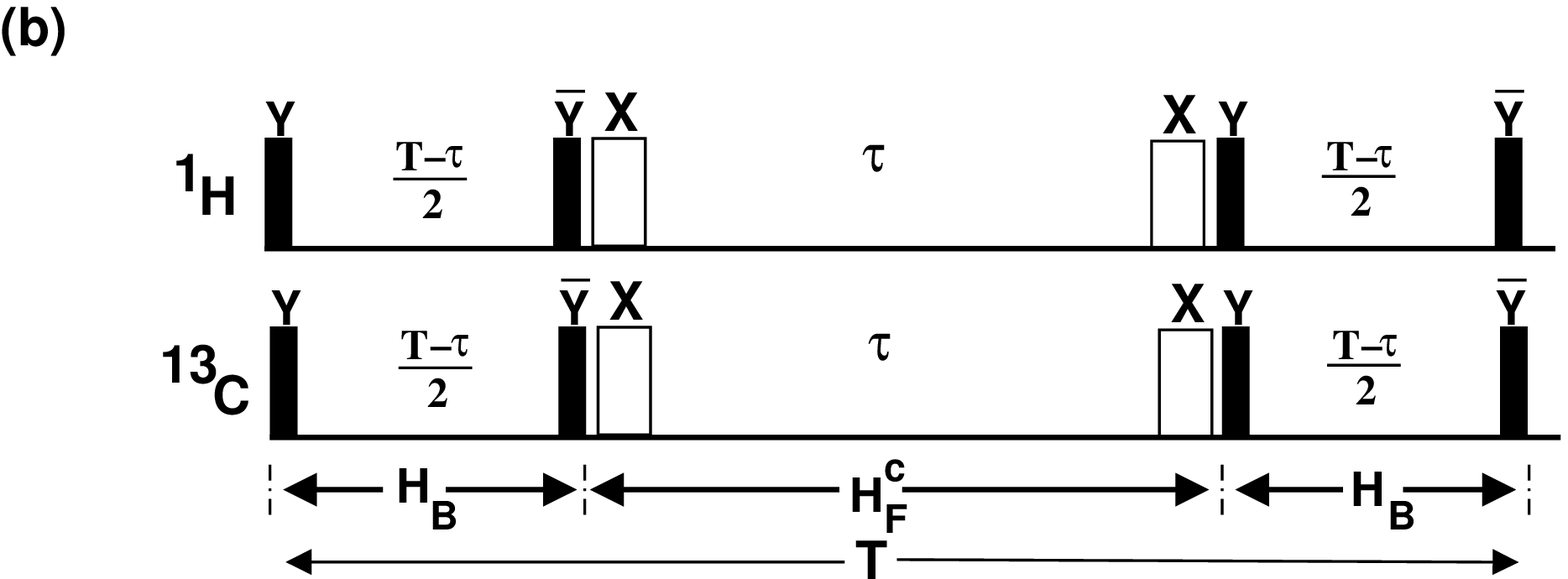,width=0.75\textwidth}}\\
\subfigure{\epsfig{file=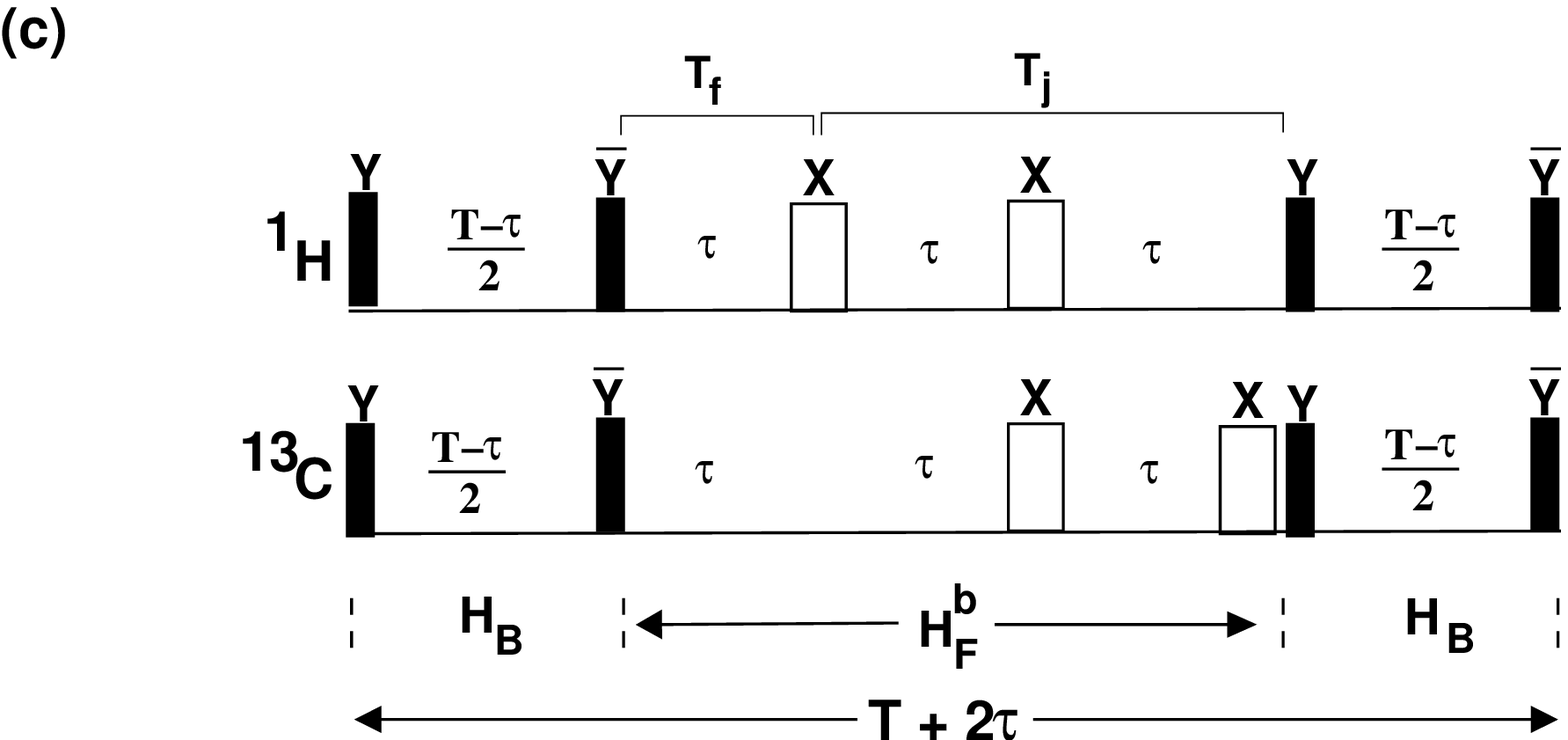,width=0.75\textwidth}}
\caption{}
\end{figure}
\end{center}

\newpage
\begin{figure}[h!]
{\bf (a)}\raisebox{-27ex}{\epsfig{file=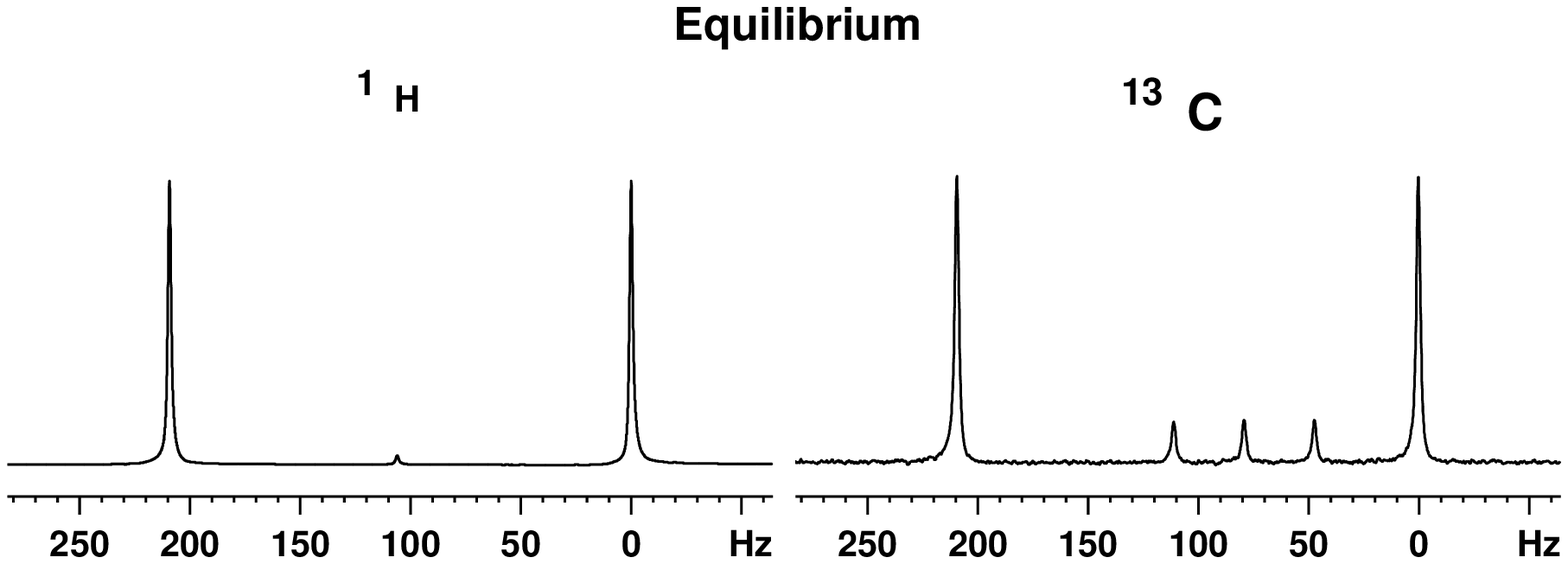,width=0.85\textwidth}}\\
{\bf (b)}\raisebox{-27ex}{\epsfig{file=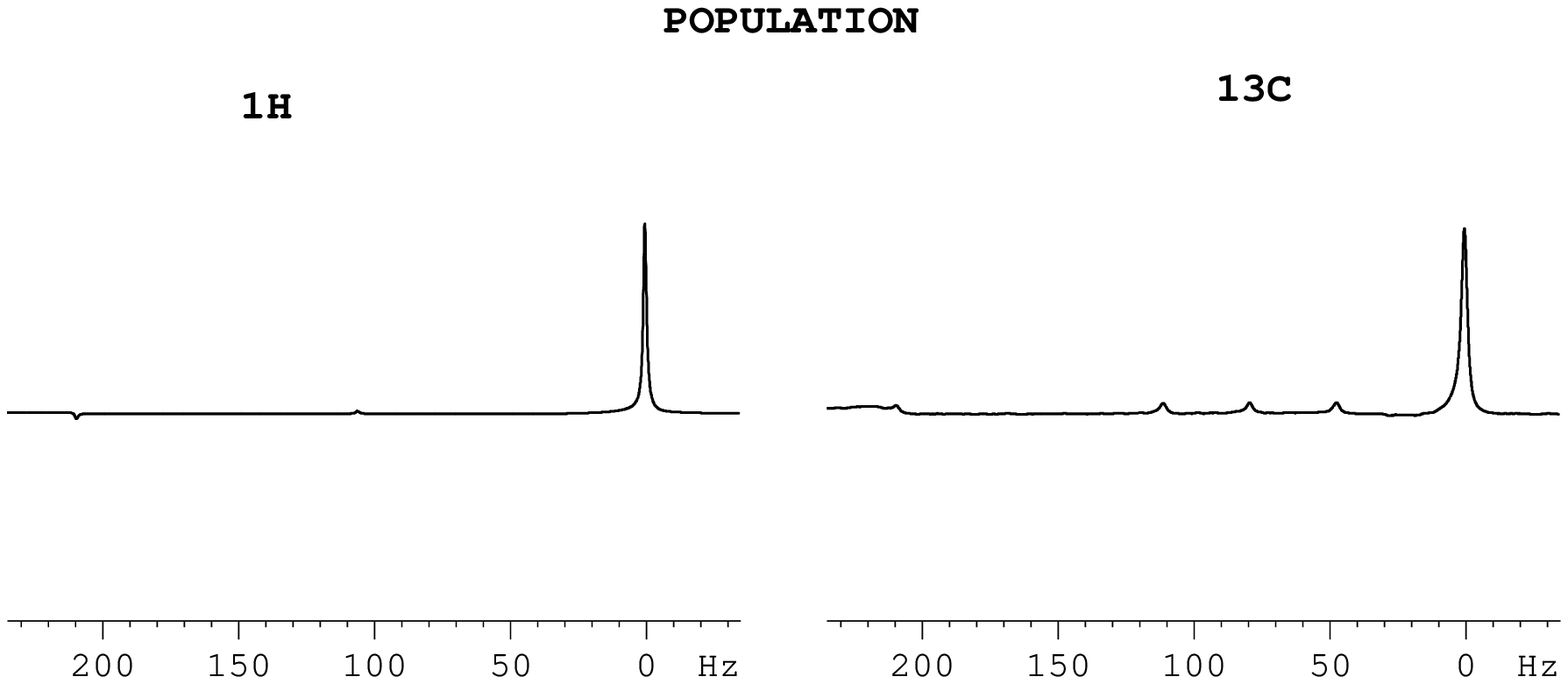,width=0.85\textwidth}}\\
\caption{}
\end{figure}

\newpage
\setcounter{figure}{0}
\renewcommand{\thefigure}{7\alph{figure}}
\begin{figure}[h!]
\subfigure{\epsfig{file=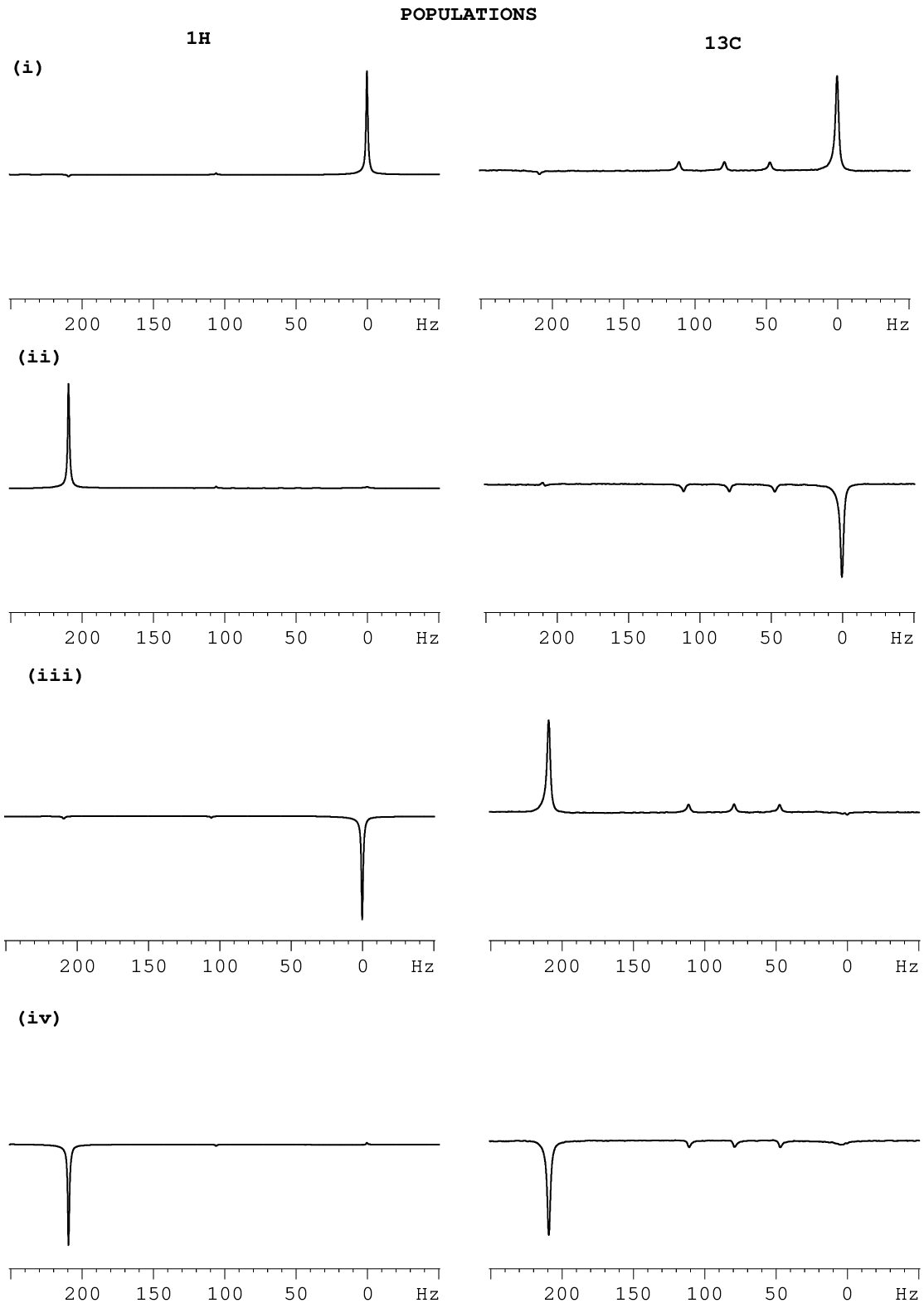,width=0.85\textwidth}}\\
\caption{}
\end{figure}

\newpage
\begin{figure}[h!]
\subfigure{\epsfig{file=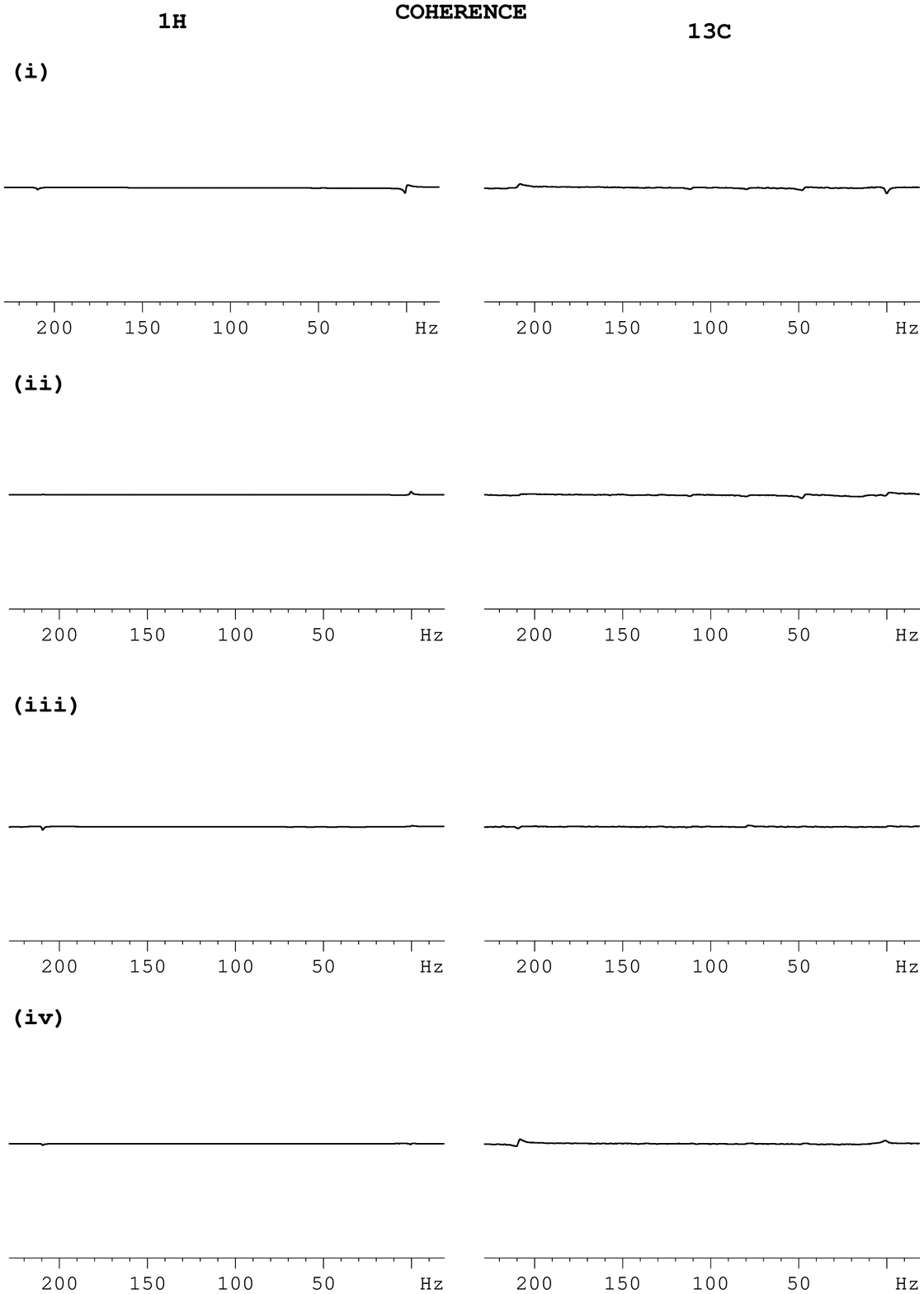,width=0.85\textwidth}}\\
\caption{}
\end{figure}

\newpage
\renewcommand{\thefigure}{8}
\begin{figure}[h!]
{\bf (a)}\subfigure{\epsfig{file=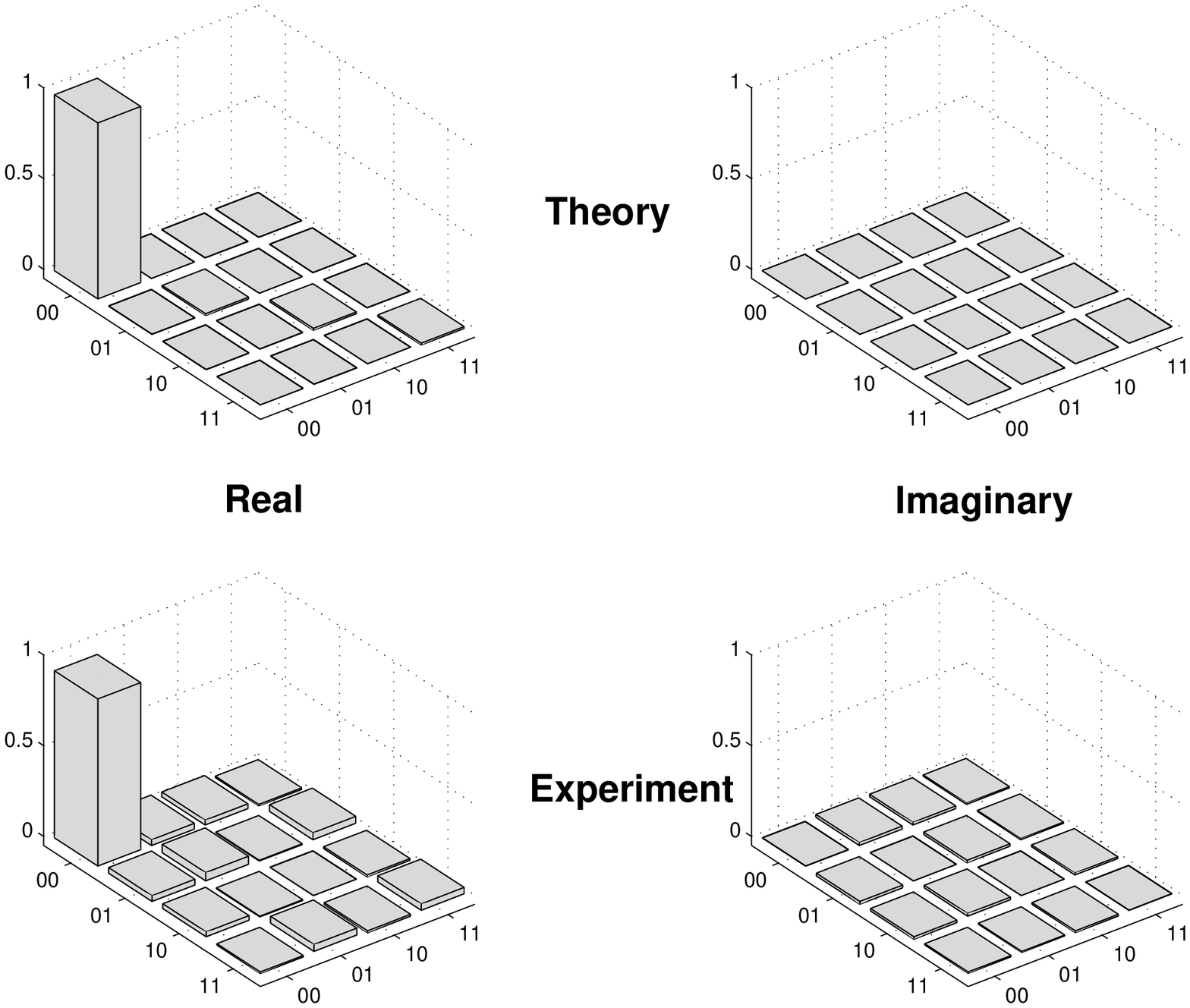,width=0.60\textwidth,angle=270}}\\
{\bf (b)}\subfigure{\epsfig{file=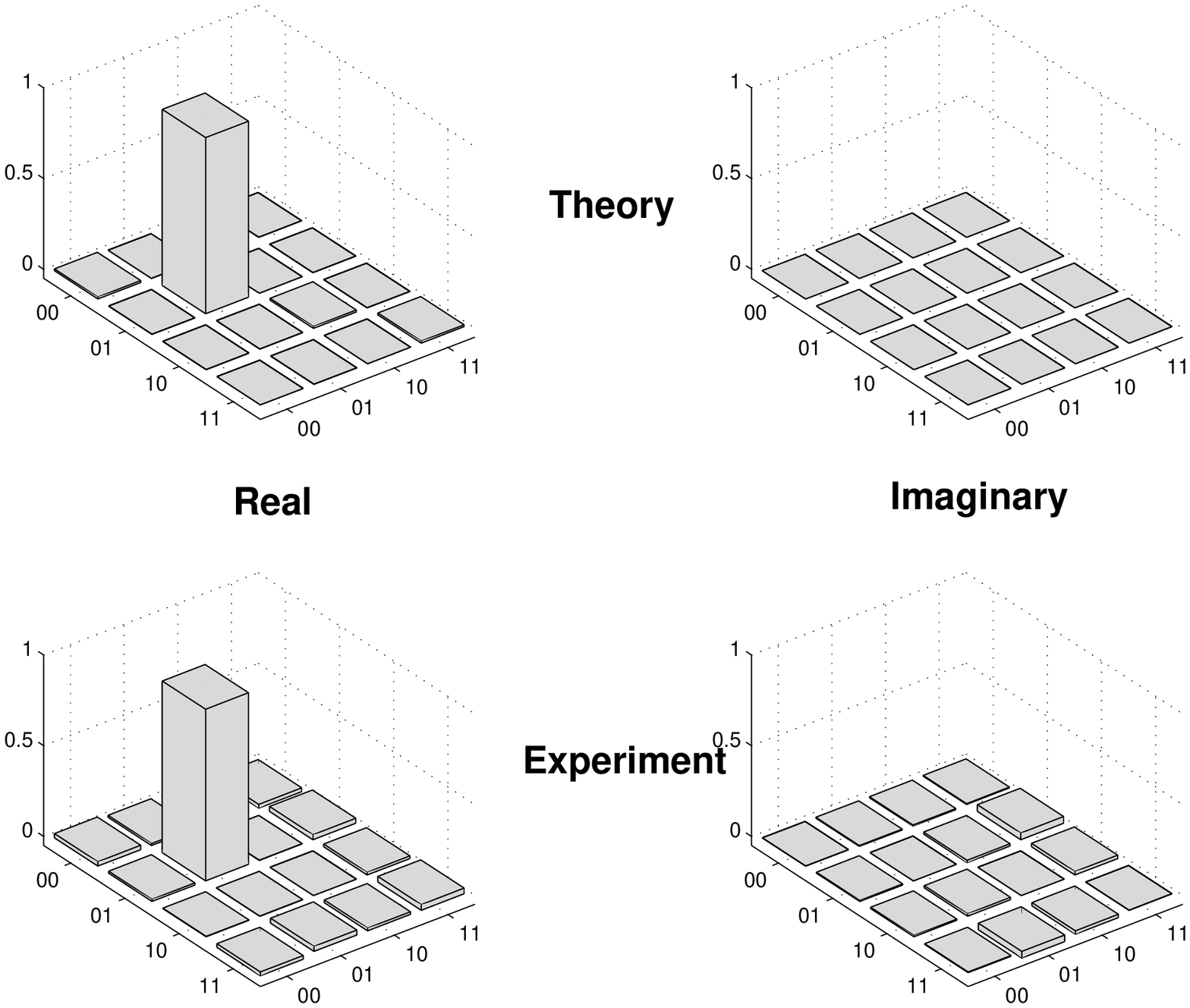,width=0.60\textwidth,angle=270}}\\
\caption{}
\end{figure}

\newpage
\renewcommand{\thefigure}{8}
\begin{figure}[h!]
{\bf (c)}\subfigure{\epsfig{file=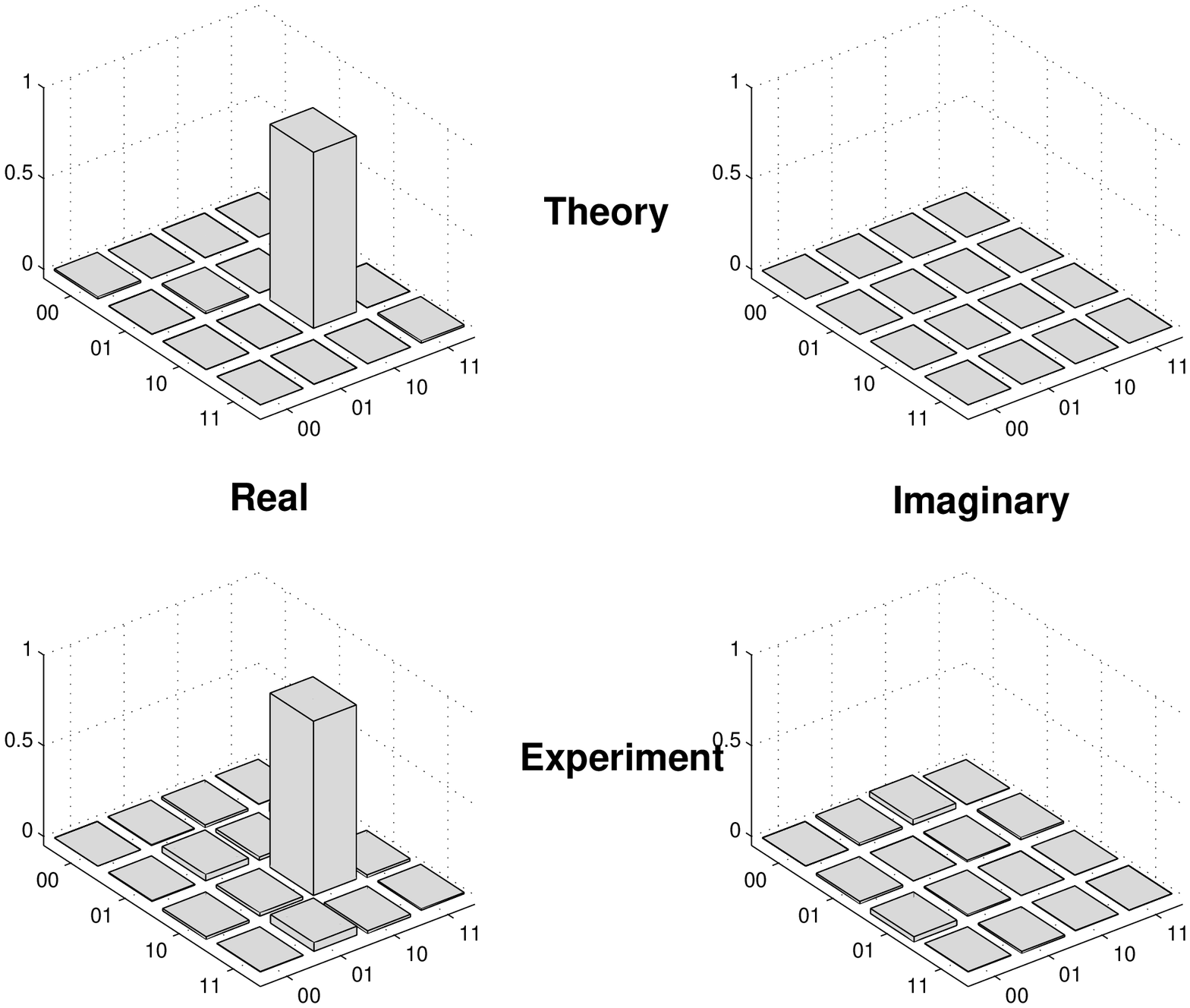,width=0.60\textwidth,angle=270}}\\
{\bf (d)}\subfigure{\epsfig{file=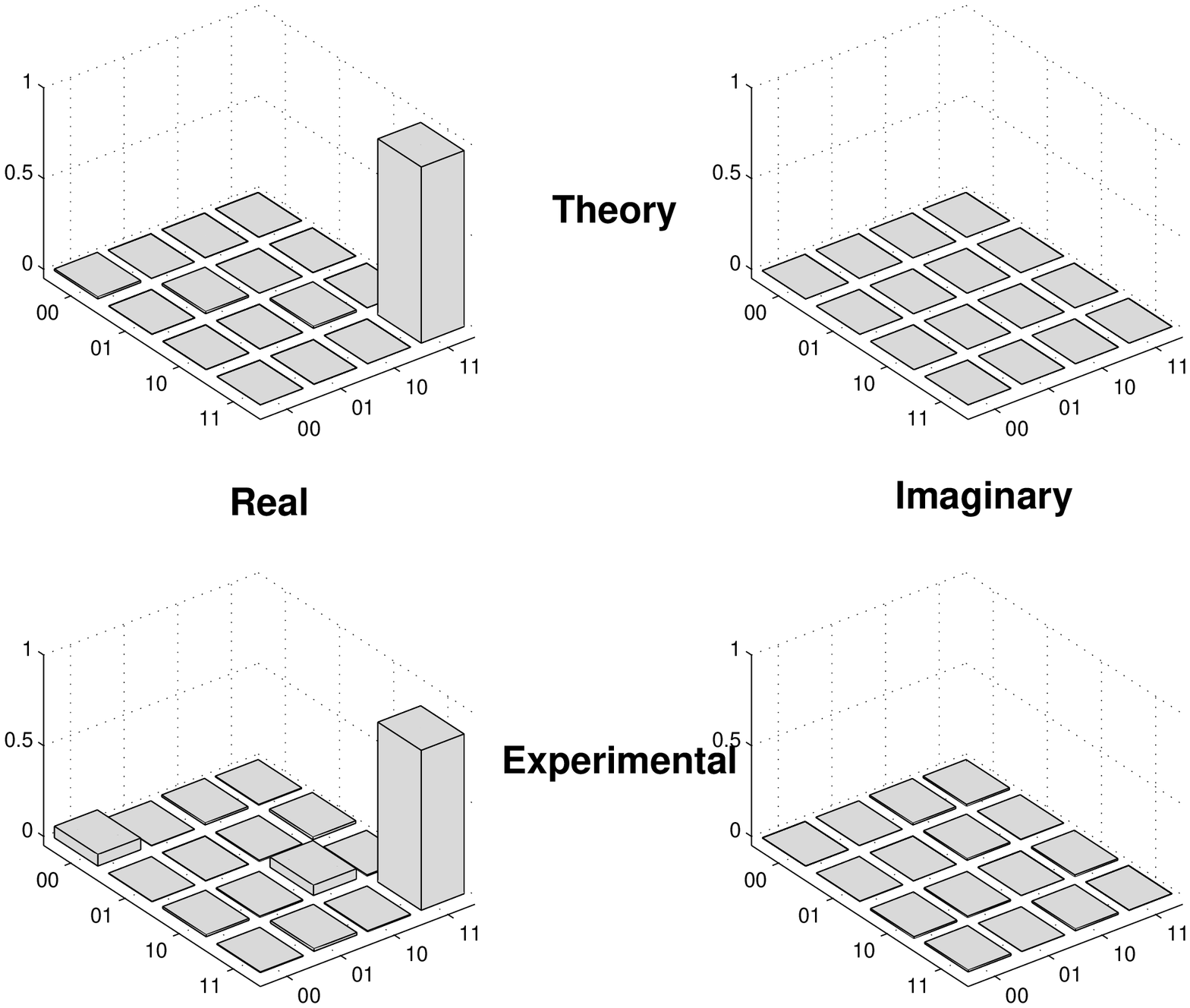,width=0.60\textwidth,angle=270}}
\caption{Cont.}
\end{figure}

\newpage
\setcounter{figure}{8}
\renewcommand{\thefigure}{\arabic{figure}}
\begin{figure}[h!]
{\bf (a)}\raisebox{-45ex}{\epsfig{file=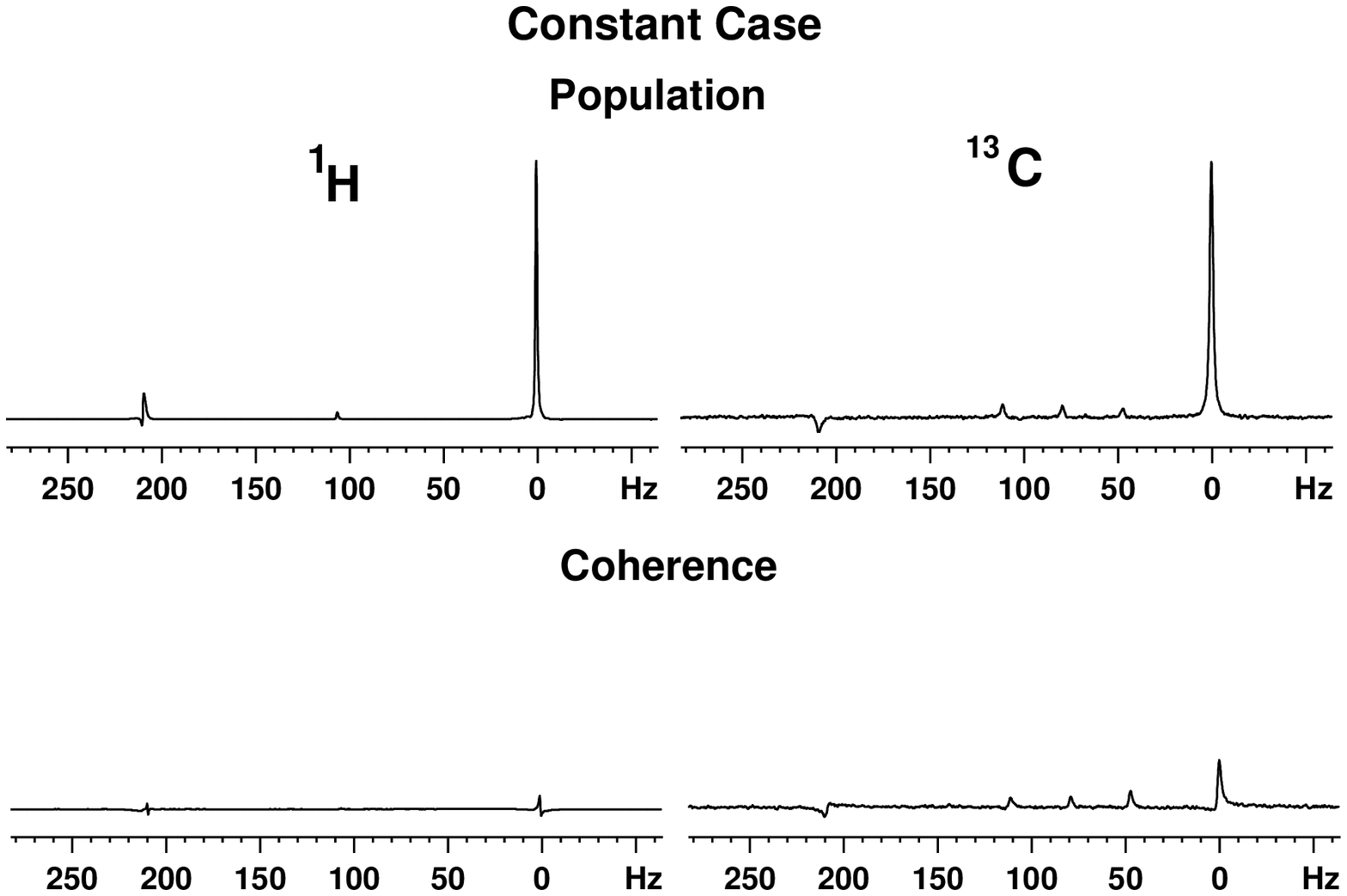,width=0.67\textwidth}}\\
{\bf (b)}\raisebox{-45ex}{\epsfig{file=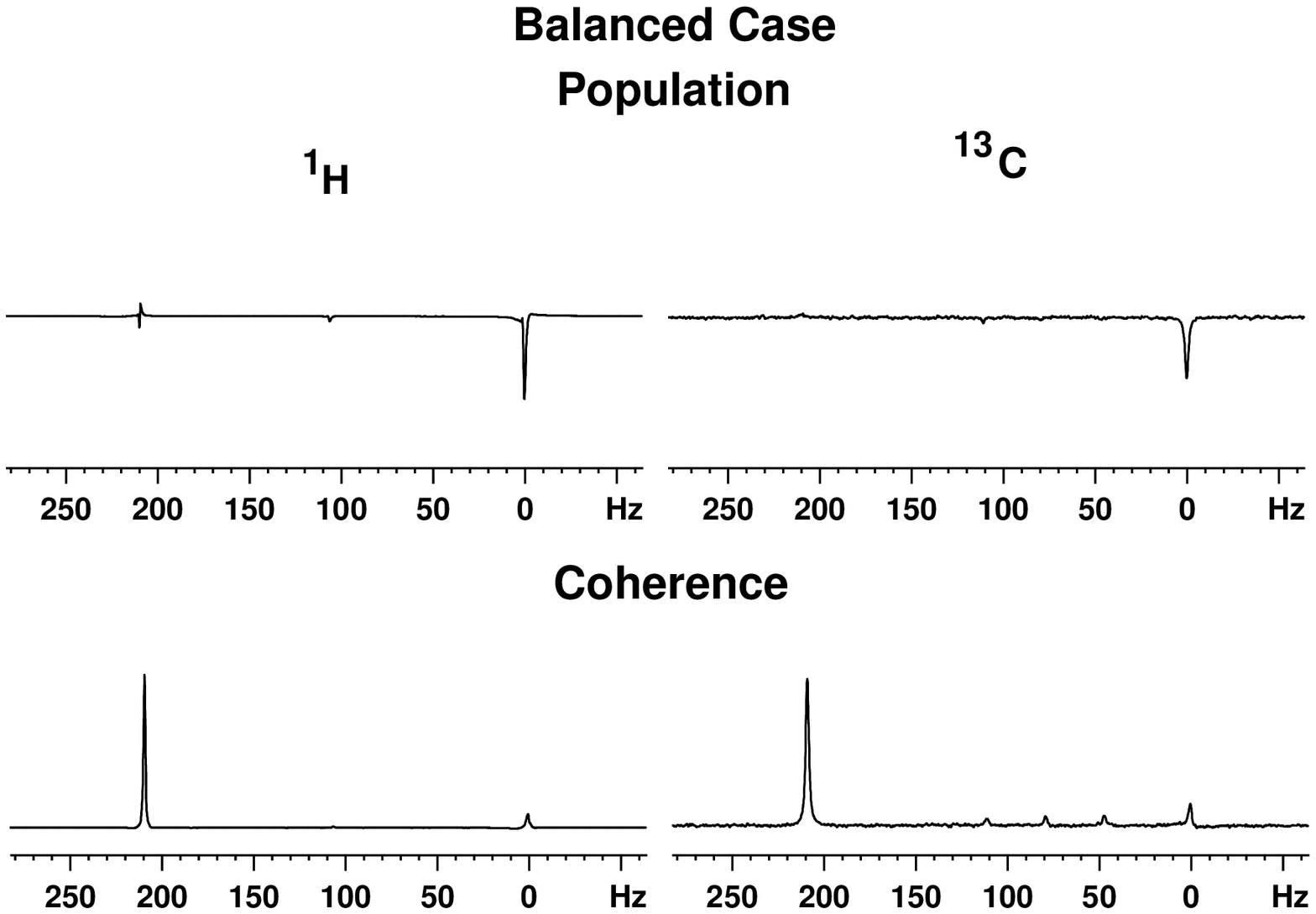,width=0.67\textwidth}}
\caption{}
\end{figure}

\newpage
\begin{figure}[h!]
\hspace*{-2cm}\epsfig{file=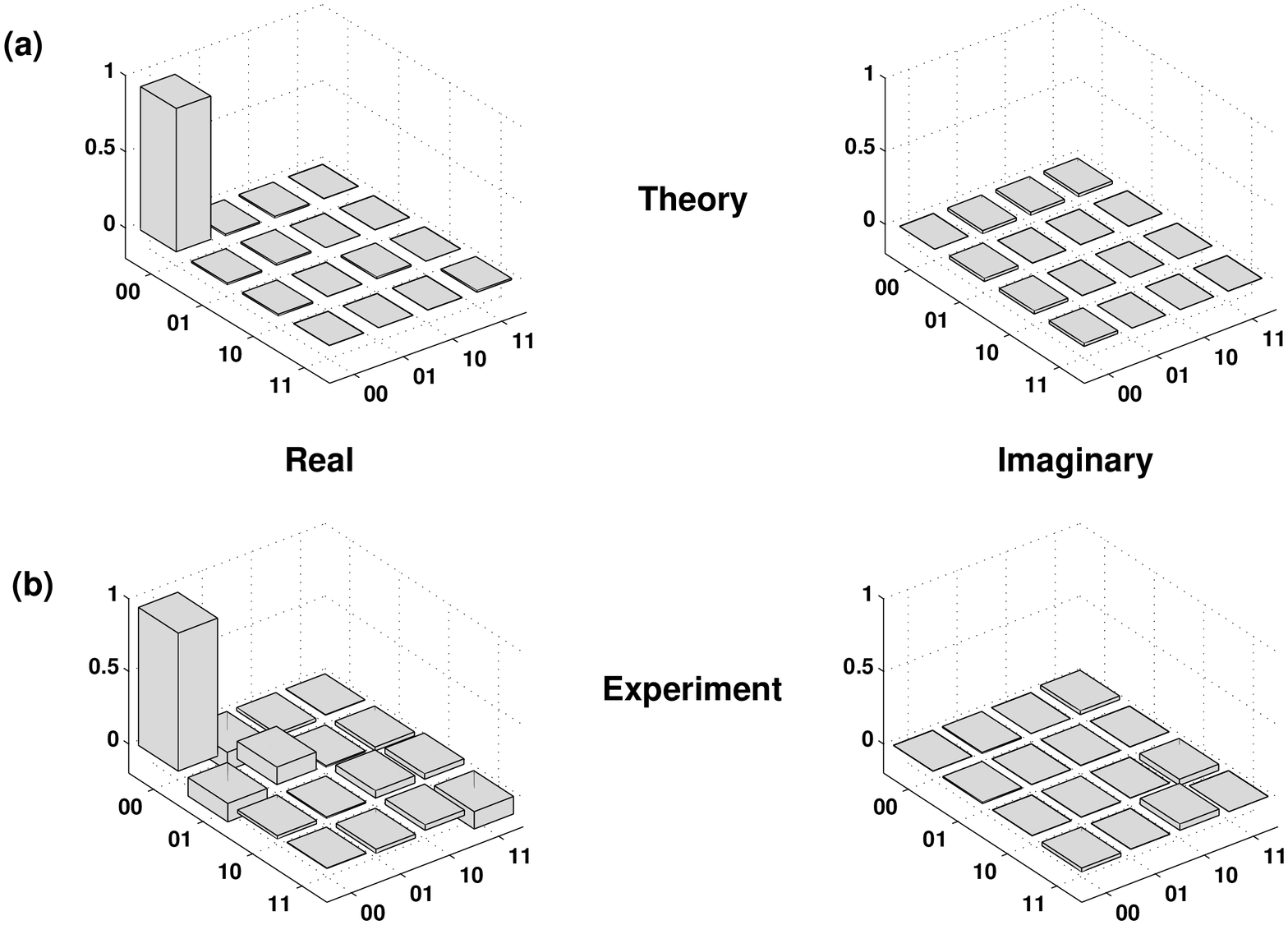,height=16cm,angle=270}
\caption{}
\end{figure}

\newpage
\begin{figure}
\hspace*{-4cm}\epsfig{file=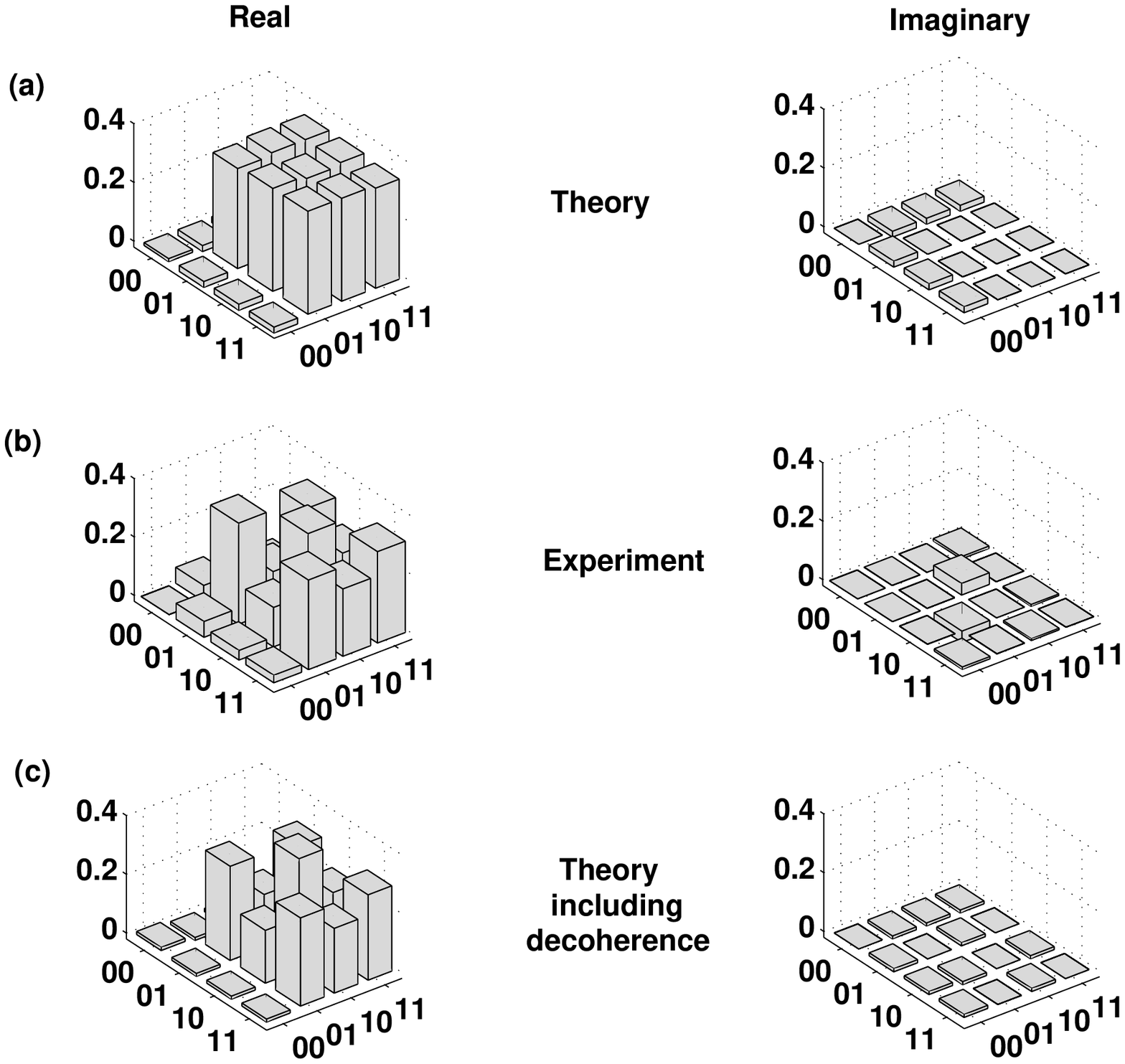,angle=270,width=1.2\textwidth}
\caption{}
\end{figure}

\end{document}